\documentclass[conference]{IEEEtran}

\usepackage{inconsolata}
\usepackage[noadjust]{cite}
\usepackage[pdftex]{graphicx}
\usepackage{amsmath}
\usepackage{algorithmic}
\usepackage{url}
\usepackage{multirow, multicol}
\usepackage{enumerate}
\usepackage[inline]{enumitem}
\usepackage{adjustbox}
\usepackage{subcaption}
\usepackage{array, booktabs}
\usepackage[ruled,vlined]{algorithm2e}
\usepackage[utf8]{inputenc}
\usepackage[bookmarks=false]{hyperref}
\usepackage{verbatim}
\usepackage{listings}
\usepackage{soul}
\usepackage{amsbsy}
\usepackage{bbding}

\usepackage{adjustbox}
\usepackage{booktabs}
\usepackage{color, colortbl}

\newcolumntype{M}[1]{>{\centering\arraybackslash}m{#1}}

\usepackage[svgnames]{xcolor}
\definecolor{diffincl}{named}{Green}
\definecolor{diffrem}{named}{Red}

\colorlet{blcolor}{gray!80}

\usepackage{tcolorbox}

\usepackage{listings}
  \lstdefinelanguage{diff}{
    basicstyle=\scriptsize\ttfamily,
    language=Python,
    frame=single,
    breaklines=true,
    morecomment=[f][\color{diffincl}]{+\ },
    morecomment=[f][\color{diffrem}]{-\ },
  }

\definecolor{Gray}{gray}{0.9}

\graphicspath{{../figures/}}
\DeclareGraphicsExtensions{.pdf,.jpeg,.png}

\def\BibTeX{{\rm B\kern-.05em{\sc i\kern-.025em b}\kern-.08em
    T\kern-.1667em\lower.7ex\hbox{E}\kern-.125emX}}

\newcommand{\lmttfont}{\fontfamily{lmtt}\selectfont}

\begin{document}
\title{Analyzing the Context of Bug-Fixing Changes in the OpenStack Cloud Computing Platform}





\author{
    \IEEEauthorblockN{Domenico Cotroneo\IEEEauthorrefmark{1}, Luigi De Simone\IEEEauthorrefmark{1}, Antonio Ken Iannillo\IEEEauthorrefmark{2}, \\Roberto Natella\IEEEauthorrefmark{1}, Stefano Rosiello\IEEEauthorrefmark{1}, Nematollah Bidokhti\IEEEauthorrefmark{3}}
    \IEEEauthorblockA{\\\IEEEauthorrefmark{1}Università degli Studi di Napoli Federico II, Naples, Italy
    \\\{cotroneo, luigi.desimone, roberto.natella, stefano.rosiello\}@unina.it}
    \IEEEauthorblockA{\IEEEauthorrefmark{2}SnT - University of Luxembourg
    \\antonioken.iannillo@uni.lu
    }
    \IEEEauthorblockA{\IEEEauthorrefmark{3}Futurewei Technologies, Inc., USA
    \\nbidokht@futurewei.com
    }
}

\maketitle
\thispagestyle{plain}
\pagestyle{plain}

\begin{abstract}
Many research areas in software engineering, such as mutation testing, automatic repair, fault localization, and fault injection, rely on empirical knowledge about recurring \emph{bug-fixing code changes}. Previous studies in this field focus on \emph{what} has been changed due to bug-fixes, such as in terms of code edit actions. However, such studies did not consider \emph{where} the bug-fix change was made (i.e., the \emph{context} of the change), but knowing about the context can potentially narrow the search space for many software engineering techniques (e.g., by focusing mutation only on specific parts of the software). Furthermore, most previous work on bug-fixing changes focused on C and Java projects, but there is little empirical evidence about Python software. Therefore, in this paper we perform a thorough empirical analysis of bug-fixing changes in three OpenStack projects, focusing on both the \emph{what} and the \emph{where} of the changes. We observed that all the recurring change patterns are not oblivious with respect to the surrounding code, but tend to occur in specific code contexts.

\end{abstract}


\section{Introduction}
\label{sec:introduction}

Studying \emph{bug-fixing changes} is an important field of software engineering research \cite{tufano2018learning, brown2017care, zhong2015empirical, soto2016deeper, koyuncu2018fixminer, lin2016empirical, musavi2016experience, osman2014mining, pan2009toward, martinez2013automatically, fluri2008discovering, duraes2006emulation, basso2009investigation, neamtiu2005understanding, silva2017refdiff, hora2018assessing}. It consists in empirically analyzing the changes made by developers to software in real complex projects, with the aim to identify (possibly, in automated ways) patterns for the most common changes, and to create profiles for these changes. This analysis is useful for many software engineering tasks, such as software testing (in particular, mutation testing \cite{tufano2018learning, brown2017care}), fault localization \cite{zhong2015empirical}, automatic code repair \cite{zhong2015empirical, koyuncu2018fixminer}, and fault injection for testing fault-tolerance \cite{duraes2006emulation, basso2009investigation}.
Analyzing bug-fixing changes can be challenging since change patterns can be numerous and heterogeneous, and they can differ across different application domains, programming languages, and even software projects.

In this paper, we analyze bug-fixing changes in the context of the OpenStack cloud computing platform \cite{OpenStackHome}. OpenStack is a widespread software, as it is adopted in many private and public cloud infrastructures \cite{OpenStackProducts} and forms the basis of more than 30 commercial products (i.e., distributions and appliances) \cite{OpenStackUsers}. One reason that makes the OpenStack platform a relevant investigation subject is that it consists in several, diverse systems that focus on different cloud computing functions (sub-systems like \textit{Nova} for managing instances, \textit{Neutron} for managing virtual networks, and \textit{Cinder} for managing volumes). These systems are developed under independent projects by separate development teams, follow rigorous development and QA processes, and have nowadays achieved a high degree of maturity \cite{openstack_user_survey}. 
Another motivation is that OpenStack is among the largest and most sophisticated software written in Python, which is a popular programming language that has not been investigated in depth by previous research on bug-fixing changes.

We first analyze in this study \emph{what} syntactic changes characterize bug-fixes in the OpenStack platform. This analysis advances the existing body of knowledge in the field of bug-fixing changes since it provides empirical insights on large Python projects. Moreover, this study explores the variability of bug-fixing changes across different projects and development teams, and discusses their variability with respect to other programming languages analyzed by previous studies (mostly on C and Java). Our approach performs a clustering analysis of bug-fixing changes, using numerical features from the Abstract Syntax Tree (AST) of the fixed code. 

In addition to syntactic changes, we also study the code context \emph{where} bug-fixing changes were made, that is, the source code that surrounds the change. Most research on code changes neglects the code context, but this aspect can potentially narrow the search space for many software engineering tasks. For example, in mutation testing and in fault injection, mutants are generated by introducing changes throughout the whole source code (for example, in the case of ``assignment omissions'', by mutating every assignment statement). However, the number of generated mutants grows very quickly or too easy to kill \cite{jia2011analysis}, with many mutants that are unkillable \cite{yao2014study}. Moreover, the size of the search space is a challenging aspect also for fault localization and for automatic code repair \cite{martinez2015mining}. Therefore, we analyze the code context surrounding the bug-fixes, to a more detailed ``fingerprint'' of the bug-fixing patterns. Our approach collects additional features to represent the context of every cluster, and points out statistical deviations that characterize the clusters.

The main findings of the study include:

\begin{itemize}[leftmargin=4mm]

    \item Commits that are supposedly bug-fixes also contain many of changes that are not strictly bug-fixes. In some cases, the changes are refactorings for supporting the bug-fix (e.g., importing a package, changing the signature and the invocations of a method, changing the layout of a data structure). In other cases, the commits are not limited to bug-fixes, but also include many changes for improving the internal quality of the software (e.g., readability and maintainability of the source code). The high number of non-bug-fixing changes points out that empirical research must take into account refactorings when analyzing bug-fixing changes for testing and repair purposes.

    \item Bug-fixing patterns exhibit relevant differences across programming languages, and across projects. While some of the bug-fixing patterns match the ones found in previous studies on C and Java software (in particular, the changes that fix the structure and the checking conditions of the control flow), we found several specific patterns that are induced by the features of the Python language, such as {\lmttfont dict} data structures and the rules for passing parameters. Moreover, we found several patterns that are specific for a project, such as, bugs influenced by API calling conventions.

    \item The bug-fixing changes tend indeed to occur in specific code contexts. For example, several change patterns were located mostly in the largest classes and methods, or were located in loops or conditional constructs. Moreover, specific traits were found for the blocks and statements impacted by the change: for example, input parameters were omitted for methods which at least 2-3 arguments, and several bug-fix patterns involved statement blocks with large numbers of data containers and function calls.

\end{itemize}

In summary, the contributions of the paper are:

\begin{itemize}[leftmargin=4mm]

\item An approach for characterizing bug-fixing changes not only with respect to \emph{what} a bug-fix changes, but also with respect to \emph{where} the change has been made.

\item A dataset of bug-fixing changes in three systems of the OpenStack platform, which represents the largest Python software ever analyzed by studies on bug-fixing changes to the best of our knowledge.

\item A detailed empirical analysis of recurring patterns in bug-fixing changes in the three OpenStack projects of the dataset.

\end{itemize}

In the following of this paper, Section~\ref{sec:related} discusses related work by exploring the various applications of bug-fix analysis; Section~\ref{sec:methodology} presents the proposed methodology for analyzing code changes; Section~\ref{sec:what_changes} and Section~\ref{sec:where_changes} analyze respectively \emph{what} is changed by bug-fixes, and \emph{where} the change was made, with a discussion on findings and implications. Section~\ref{sec:threats_to_validity} discusses the threats to validity of this study. Section~\ref{sec:conclusion} concludes the paper.

\section{Related work}
\label{sec:related}


Several studies have been investigating bug characteristics for various software engineering tasks, by analyzing problem reports, commits, revisions, and other information. Our intention is to give a broad view of how researchers have been using bug information in a different context, and their main findings. 
In Appendix A, we summarize the surveyed studies, highlighting the specific \textit{purpose}, the \textit{programming language} used in the software under study, the \textit{dataset} (e.g., number of code changes, revisions, commits, bug-fixes), and a brief description of the \textit{findings} related to  bug-fix patterns.

In most of these studies, the authors analyze code changes by using an \textit{Abstraction Syntax Tree (AST)} (i.e., a data structure representation of entities in the source code and their relations), and a generate \textit{edit actions} that reflect the differences between the ASTs before and after a change of the source code.


\textbf{\textit{Mutation testing}}. Mutation testing is a fault-based technique for the creation and the assessment of test suites. 
Recently, Tufano et al. \cite{tufano2018learning} developed an AST-based differencing technique for analyzing bug-fixes and to abstract them. Their approach trains an artificial neural network with the bug-fixes, and then use the neural network to introduce new mutants that reflect the learned ones. 
Brown et al. \cite{brown2017care} introduced the concept of \textit{wild-caught-mutants}, to address the issues that mutation operators do not necessarily emulate the types of changes made to source code by human programmers. Thus, their idea is to analyze bug-fixes from bug reports to define mutation operators that more closely reflect faults occurred in a specific project. For example, the authors found new mutation operators like missing call to a one-argument function whose return type is equal to its argument's type, direct access of field, and specific literal replacements.



\textbf{\textit{Automatic program repair}}. Automatic program repair is a branch of research on lowering the costs of bug-fixing. The general approach is to locate and mutate a faulty source location with a set of change operators, using search-based techniques, until the program passes a test suite. The quality of the test suite and of the program under fixing are preconditions for generating patches with a reasonable chance of success. 
Zhong et al. \cite{zhong2015empirical} performed an empirical study on fixes of real bugs in open-source projects in order to reuse change patterns for automatic repairing and understand to what extent bugs are localized. 
Similarly, Koyuncy et al. \cite{koyuncu2018fixminer} implemented repair strategies based on fix patterns or templates. They provide a tool for mining semantically-relevant patterns in a scalable, accurate and actionable way, by using a clustering strategy.


\textbf{\textit{Refactoring}}. Bug characterization studies analyzed whether an issue marked as a bug is actually a bug. As a matter of fact, in a recent study Herzig et al. \cite{herzig2013s} found that a high number of non-bug issue reports are misclassified as bugs, such as refactorings, requests for new features, documentation, and test cases.
In particular, previous studies on refactoring use source code changes history to detect and study refactoring changes. 
%
The approach by Silva et al. \cite{silva2017refdiff} consists in 2 phases: (i) parse and analyze the history of source code changes to obtain a high level abstraction (i.e., a multiset of tokens); (ii) perform a relationship analysis, i.e., the procedure to find similarities between source code abstractions before and after the changes. The method was able to find 12 well-known refactoring templates (as defined by Prete et al. \cite{prete2010template}) with a \textit{Precision} of 1.00 and a \textit{Recall} of 0.88. 
Hora et al. \cite{hora2018assessing} analyzed refactorings due to so-called \textit{untracked changes}, e.g., a method rename or move. That change can be misinterpreted as the disappearance of a method and the appearance of a brand new one, splitting its history, and could have a negative impact on the accuracy of mining software repository techniques if not properly handled. 
In our work, we observe that refactoring-related changes after often mixed with bug-fixing changes, and we discriminate between these two categories to focus on bug-fixing ones.


\textbf{\textit{Fault injection}}. Fault injection is a technique for experimental evaluation of fault tolerance mechanisms, such as for quantifying their coverage and latency \cite{NFV_bench}. One research branch in this area has been focusing on the injection of software faults using code mutations, to emulate the most common bug patterns \cite{natella2013fault}. 
To ensure the representativeness of the injected bugs with respect to actual bugs, these approaches have been relying on the analysis of bug-fixing patterns. For example, previous studies \cite{duraes2006emulation, basso2009investigation} manually analyzed bug fixes in C and Java projects, with respect to an extended version of the Orthogonal Defect Classification (ODC) schema \cite{chillarege1996orthogonal}, by including in the classification the specific kind of omitted or wrong construct (assignment, control flow checking, etc.), and an early notion of ``context'' (e.g., number of statements inside an IF block, presence of an assignment before a function call, etc.). These studies identified consistent patterns across these languages, such as missing function calls and missing IF blocks; moreover, they found that one recurring bug patterns (i.e., \textit{Assignment}) are common across projects and cover 21.4\% of the total bugs, but the remaining share of bugs follows project specific patterns. 
Our study of bug-fixing patterns can be leveraged for injecting bugs into Python software and enriches the classification of bugs with broader and quantitative information about the context of bugs.



\textbf{\textit{Bug characterization}}. Numerous other studies have been investigating bug-fixing patterns, beyond the specific tasks above. 
Pan et al. \cite{pan2009toward} found that the most common categories of bug-fix patterns in Java projects are \textit{Method Call} (21.9-33.1\%) and \textit{If-Related} (19.7-33.9\%). In particular, within the Method Call category, most of the bug fixes to method calls are changes to the actual parameter expressions (14.9–25.5\%), and within If-Related category the \textit{change in if conditional} is the more frequent (5.6–18.6\%). They also provided evidence of similarities of bug-fix patterns across different projects (i.e., Pearson similarity measures exceed $0.85$ with p-value less than $0.001$), and pointed out that developers can introduce the same kind of bugs independently from the specific program domain.
Osman et al. \cite{osman2014mining} presented another analysis of Java projects, and found that 53\% of the fixes involve only one line of code, and that 73\% of fixes consist in less than 4 lines of code. Moreover, they found that 40\% of bug-fixes are recurrent patterns.  The most frequent fix pattern (48\%) involves the addition of \emph{null} checks on Java object references. Other fix patterns involve missing method invocations and wrong names for objects, methods, or parameters.
Other studies \cite{fluri2008discovering, martinez2013automatically}, pointed out similar findings.

Only a minority of studies focused on the Python language. Lin et al. \cite{lin2016empirical} analyzed 10 Python projects. They developed a tool for analyzing Python source code, and classifying changes according to edit actions on ASTs. They analyzed the distribution of edit actions across 8 gross categories, including \emph{Class} edits, \emph{Function} edits, \emph{Statement} edits, etc.. In most of the projects, they found that \textit{Function} and \textit{Statement} edits are the most common change types, whereas \textit{Loop Structure} edits are the least common ones. Furthermore, the authors found that the majority of bug-fix edit actions are \textit{Conditional Expression Update} and \textit{If Insert}. 
Musavi et al. \cite{musavi2016experience} conducted an empirical study to understand API failures in OpenStack, by analyzing the code change history. The authors manually evaluate the bug reports and bug fixes of API failures during 2014, and classified them into 7 categories. More than half (56\%) causes of API failures are ``small programming faults'', which were fixed with simple edits such as inverting logical conditions, correcting variable names, or adding exception handling. 
%

The main points of difference between our work and these studies can be summarized as follows:

\begin{itemize}[leftmargin=4mm]

    \item Most of the previous studies focused on software projects written in Java, for which there exist more various and mature tools for analyzing source code characteristics. Instead, our work concentrated on the less-explored, but much relevant Python language. We provided new insights about recurrent fix patterns found in large projects written in Python. Compared to the few previous studies on Python, we performed a more fine-grained analysis of bug-fixing patterns, not limited to distributions of changes with respect to fixed categories (e.g., type of edit actions or small-vs-large programming faults) but using clustering to discover patterns in an unsupervised way.

    \item Recent research on bug patterns did not focus on the context in which code changes were made. Almost all previous studies have discovered that some bug patterns are more frequent than others, but do not give enough information about ``where'' in the code the bug occurred. Therefore, our analysis provides more detailed insights on the context of bug-fixes.
    
\end{itemize}






\section{Methodology}
\label{sec:methodology}

\begin{figure}
    \centering
  \includegraphics[width=0.50\textwidth]{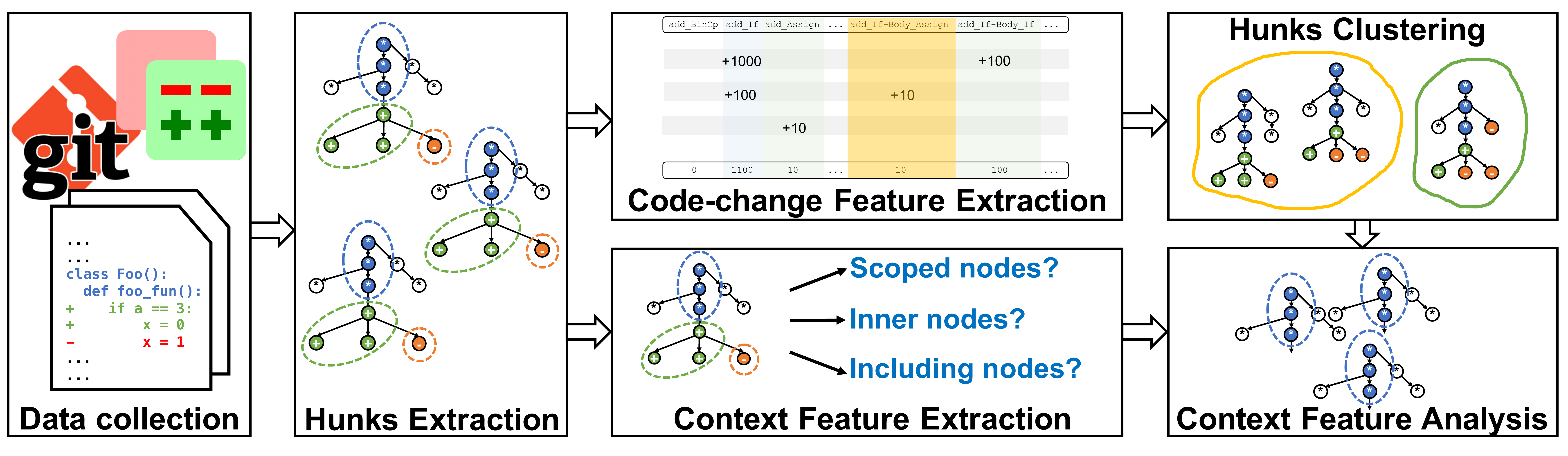}
  \caption{The proposed approach.}
  \label{fig:approach}
\end{figure}

The proposed approach consists of the following phases, which are summarized in \figurename~\ref{fig:approach}. First, we harvest data from the OpenStack public repository (\subsectionautorefname~\ref{subsec:data_collection}), by collecting \emph{code changes}. Then, we extract \emph{hunks} (see \tablename~\ref{tab:definition} about the terminology used by the OpenStack project) for the source files involved in the change (\subsectionautorefname~\ref{subsec:hunk_extraction}), generate features from these hunks (\subsectionautorefname~\ref{subsec:code_change_feature_extraction}), and we perform clustering of these hunk according to their features (\subsectionautorefname~\ref{subsec:hunk_clusterization}). The clusters will represent recurring patterns for bug-fixes. Once the clusters are defined, we investigate the code context surrounding the hunks, by means of an inter-cluster analysis on an additional set features (\subsectionautorefname~\ref{subsec:context_feature_analysis}). The resulting dataset, which include the code-change and the context features extracted from Openstack public repository, the results of both bug-fix clustering and the context analysis, is publicly available online at \url{https://figshare.com/s/7ae9d7dade9e8df62683}.

\begin{table}
\centering
\caption{Definitions.}
\label{tab:definition}
\begin{tabular}{m{.7cm}m{2cm}m{4.8cm}}
\toprule
\textbf{Name} & \textbf{Definition} & \textbf{Description} \\ \midrule

Change & A set of patches with comments and code review rating & When developers want to fix a new bug or add new functionality, they push a commit with a new id (Change-Id) and Gerrit create a change. \\\midrule

Revision & A newer version of the change & A change has an initial version and possible multiple following versions. When developers want to modify their change, they push a new commit with the same Change-Id of the initial commit, and Gerrit creates a new revision to the change by adding the new set of patches and allowing new comments and rating. \\\midrule

Merged Change & A change which has been accepted by reviewers & When a change is accepted by reviewers, Gerrit cherry-pick the last revision's patches into the repository’s master branch and mark the change as merged. \\\midrule

Hunk & A group of consecutive lines that were modified by a patch & A hunk includes both the lines of the source code before the change and the lines of the source code after the change. \\ \bottomrule

\end{tabular}
\end{table}

\subsection{Data collection}\label{subsec:data_collection}

The source of data that we analyze in this study comes from \textit{Gerrit}, the code review system used by the OpenStack project \cite{openstack_gerrit}, which is openly accessible. 
We query the repository to collect the latest \textit{revision}s of each \textit{merged change}s and the list of files modified by the revision's commit. 
Then, we filter the query results to focus on bug-related changes. We analyze the description of the change, and we only retain the changes that include at least one of the following keywords: \emph{bug}, \emph{fix}, \emph{fault}, \emph{fail}, \emph{patch}. A similar approach has been already adopted in other studies \cite{kim2008classifying, kamei2013large, musavi2016experience}.
Revisions of changes that do not contain any of these keywords are discarded. We also discard those files that only contain test cases because they represent unnecessary information for our analysis, as we are only interested in the specific patches needed to fix the bugs.

Our analysis focuses on the data related to the OpenStack Nova, Neutron, and Cinder sub-projects, respectively the compute, network, and storage managers of the OpenStack platform. Furthermore, we focus on the last four versions of OpenStack, \emph{i.e.,} Ocata, Pike, Queens, Rocket releases. 
In total, we collected \textbf{22,418 unique revisions}, which touch \textbf{45,428 files}. The time span of the revisions is from February 2017 to May 2018.


\subsection{Hunks Extraction}
\label{subsec:hunk_extraction}

We iteratively analyze the collected files to extract  \textit{hunk}s. First of all, the data from Gerrit contain only the git references to the actual files and they are retrieved automatically during this step of the analysis. Furthermore, they are converted to an Abstract Syntax Tree (AST) for convenience of manipulation and analysis of source code.

For each file, we parse the two versions (before and after the fix) of the source code to their respective ASTs. Then, we join the two trees to create an enhanced AST (\textit{AST of differences}). In such a tree, a node can be labeled as \textbf{minus node}, i.e., a node removed to fix the bug, or \textbf{plus node}, i.e., a node added to fix the bug. The nodes that are not labeled represent the parts of the source file that were not modified by the change. 
The plus and minus nodes are grouped in hunks, such that nodes in the same hunk are within three lines of the source code, as bug-fixing changes tend to focus on localized portions of source code \cite{osman2014mining, musavi2016experience}.

Since the hunk includes a subset of nodes, it represents a sub-tree of the enhanced AST. The hunk can be a single node (\emph{e.g.,} when the bug-fix just changes the name of a variable) or a whole sub-tree (\emph{e.g.,} the bug-fix changes an \emph{if} block that contains several statements). A hunk is also characterized by all the ancestors of its plus and minus nodes. These ancestors are unlabeled nodes, which we define as \textbf{context nodes}. These nodes tell us which are the source code that envelops the change. For example, context nodes give information whether the changed lines are inside constructs like \emph{if}, \emph{for}, \emph{with}, \emph{while}, \emph{function definition}, \emph{class definition}, or a combination of them. 

\begin{figure}
  \centering
  \includegraphics[width=.8\columnwidth]{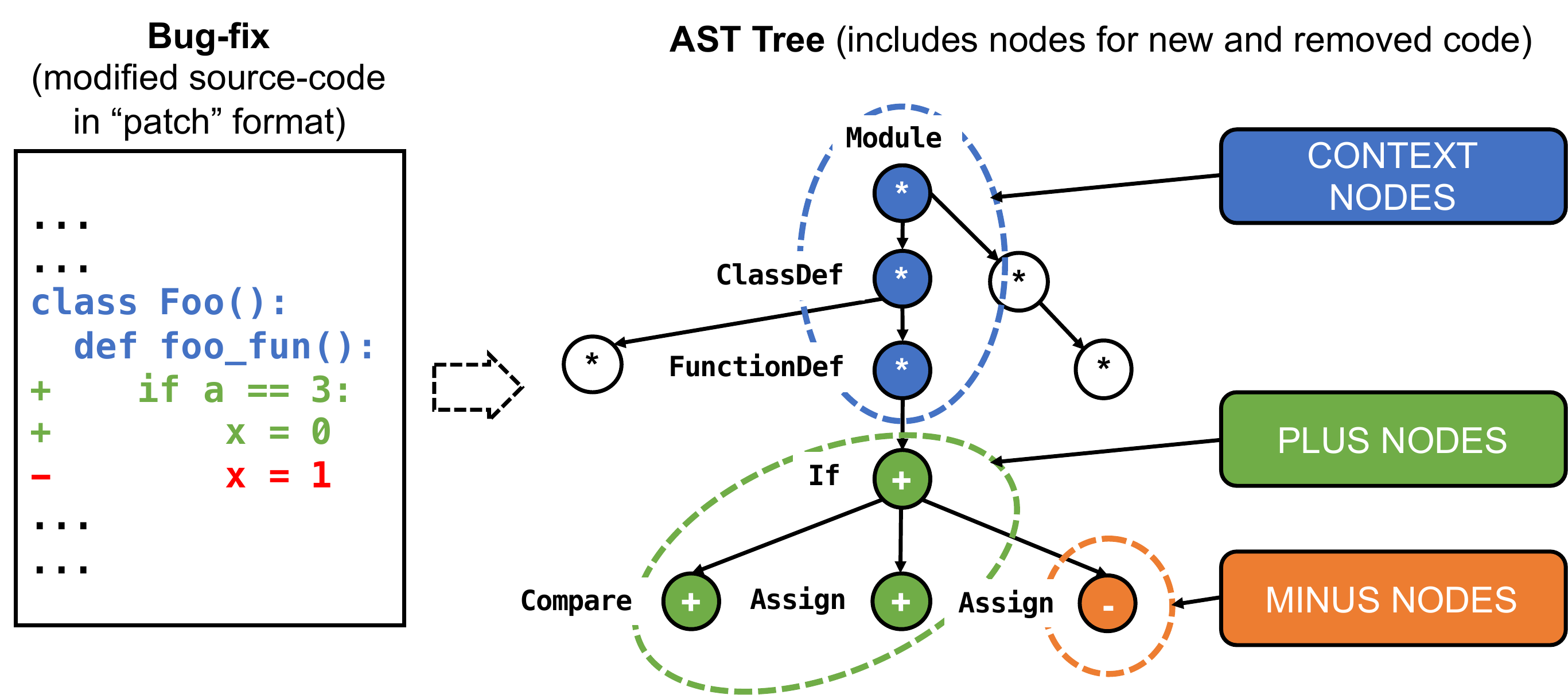}
  \caption{Example of enhanced AST.}
  \label{fig:enchanced_ast_example}
\end{figure}

\figurename~\ref{fig:enchanced_ast_example} shows an example of an enhanced AST tree for a change. Specifically, the enhanced AST tree represents a change made within a function named {\lmttfont foo\_fun} defined inside the class {\lmttfont Foo}, by adding an if construct whose body has been replaced the initialization of a variable (i.e., {\lmttfont x = 0}). 

We designed and developed a tool for fully automate this step of the analysis, namely \textit{PySA} (Python Source-code Analyzer) (publicly available at \url{https://github.com/dessertlab/PySA2}). \emph{PySA} is able to: (i) create a AST of differences from two versions of a file and (ii) extract the hunks from an AST of differences. We remark that we do not consider other existing AST differencing tools (e.g., ChangeDistiller \cite{fluri2007change}, GumTree \cite{falleri2014fine}) because they are either designed to work with Java or C source code, or they are not publicly available. Furthermore, such tools do not provide any information about the context, which is fundamental for our analysis.

In total, we extracted \textbf{16,081 unique hunks}, where 5,890 are from Nova, 4,261 from Neutron, and 5,930 from Cinder.

\subsection{Code-Change Features Extraction}
\label{subsec:code_change_feature_extraction}

From each hunk, we generate a \emph{feature vector} that describes the hunk as a flat series of numerical attributes. The feature vector still takes into account the relationship between statements (\emph{e.g.,} whether a statement is inside another block of code) by using \emph{weights}. The features are built by traversing the AST sub-tree for the hunk and inspecting the attributes of its nodes.

The Python Abstract Grammar consists of \textbf{89 AST node types} (e.g., an {\lmttfont If} node, a {\lmttfont Call} node, etc.). Moreover, an AST node can take over one of the \textbf{96 AST roles}, depending on the type of AST node. For example, a Python expression (represented by the {\lmttfont Expr} AST node type) can appear inside an {\lmttfont If} block, thus taking the role {\lmttfont If-Body}; or, a Python expression can appear as an input parameter of a method call, thus taking the role {\lmttfont Call-Args}. 

We define a \textbf{feature vector} in which each element specifies (i) whether a node was added (i.e., \emph{plus} nodes) or removed (i.e., \emph{minus} nodes) within the fix, and (ii) counts how many times a node belongs to a specific type or role. In particular, the feature vector consists of two parts:

\begin{itemize}[leftmargin=4mm]

    \item \textbf{Node type features}: For each AST node type (e.g., \emph{Assign}, \emph{Call}, etc.), we have a feature that keeps track of how many times that node type appears in the bug-fix. In total, there are \textit{178 node type features}, defined as {\lmttfont{<add|rem>\_<node\_type>}};
    
    \item \textbf{Role type features}: For each AST node type, we have 96 potential \textit{role types}, which specifies the relationship between a AST node and its parent. Thus, these features keep track how many times an AST node type has a specific role. In total, there are \textit{17,088 role type features}, defined as {\lmttfont <add|rem>\_<role>\_<node\_type>}.

\end{itemize} 


Starting from the enhanced AST, we check the type and role of each node, then we increase accordingly the corresponding element in the feature vector. Each type/role element is increased for every occurrence of that type/role in the AST. Specifically, the element is increased by a \emph{weighted value}, which takes into account the depth in which the type/role appears in the AST tree. This allows us to preserve part of the information about the structure of the code in the hunk.
In particular, the node type feature $F_{type}$ for a node type $type$ is increased by:

\begin{equation}
    \boldsymbol{F}_{type} \mathrel{+}= \boldsymbol{w}_{type} \times \boldsymbol{r}^{\boldsymbol{-level}}
\end{equation}


\noindent
for each node of that type in the hunk, where:
\begin{itemize}[leftmargin=4mm]
    \item $\boldsymbol{w}_{type}$ is a weight that represents the relative importance between AST node types;
    \item $\boldsymbol{r}$ is the relative importance between a node and its parent;
    \item $\boldsymbol{level}$ is the distance of the node from the root of the hunk AST tree.
\end{itemize}

In our approach, we give the same importance to all node types. In particular, we set $\boldsymbol{w}_{type} = 10^{15}$ because 15 is the maximum height a hunk AST tree have in our datasets. Thus, we force the feature to be integers. We set ${r}=10$, so that nodes at different depths are differentiated by different orders of magnitude of the counter.

In a similar way, the role type feature $F_{role}$ for a node role $role$ is increased by:

\begin{equation}
    \boldsymbol{F}_{role} \mathrel{+}= \boldsymbol{w}_{role} \times  \boldsymbol{r}^{\boldsymbol{-level}} \times \boldsymbol{c}
\end{equation}

\noindent
for each node of that role type in the hunk, where:
\begin{itemize}[leftmargin=4mm]
    \item $\boldsymbol{w}_{role}$ is a weight that represents the relative importance between AST role types;
    \item $\boldsymbol{r}$ is the relative importance between a node and its parent;
    \item $\boldsymbol{level}$ is the distance of the node from the root of the hunk AST tree;
    \item $\boldsymbol{c}$ is the relative importance between the node type and role type features.
\end{itemize}


In our approach, we give the same importance to all role types. Again, we set $\boldsymbol{w}_{role} = 10^{15}$ because 15 is the maximum height a hunk AST tree have in our datasets. Thus, we force the feature to be integers. Again, we set ${r}=10$. Furthermore, we set $\boldsymbol{c}=10^{-1}$ so that the features related to roles have a lower weight compared to node types. This choice is made in order to give greater importance to the outer code in a block of bug-fix statements.

\begin{figure}
  \centering
  \includegraphics[width=0.85\linewidth]{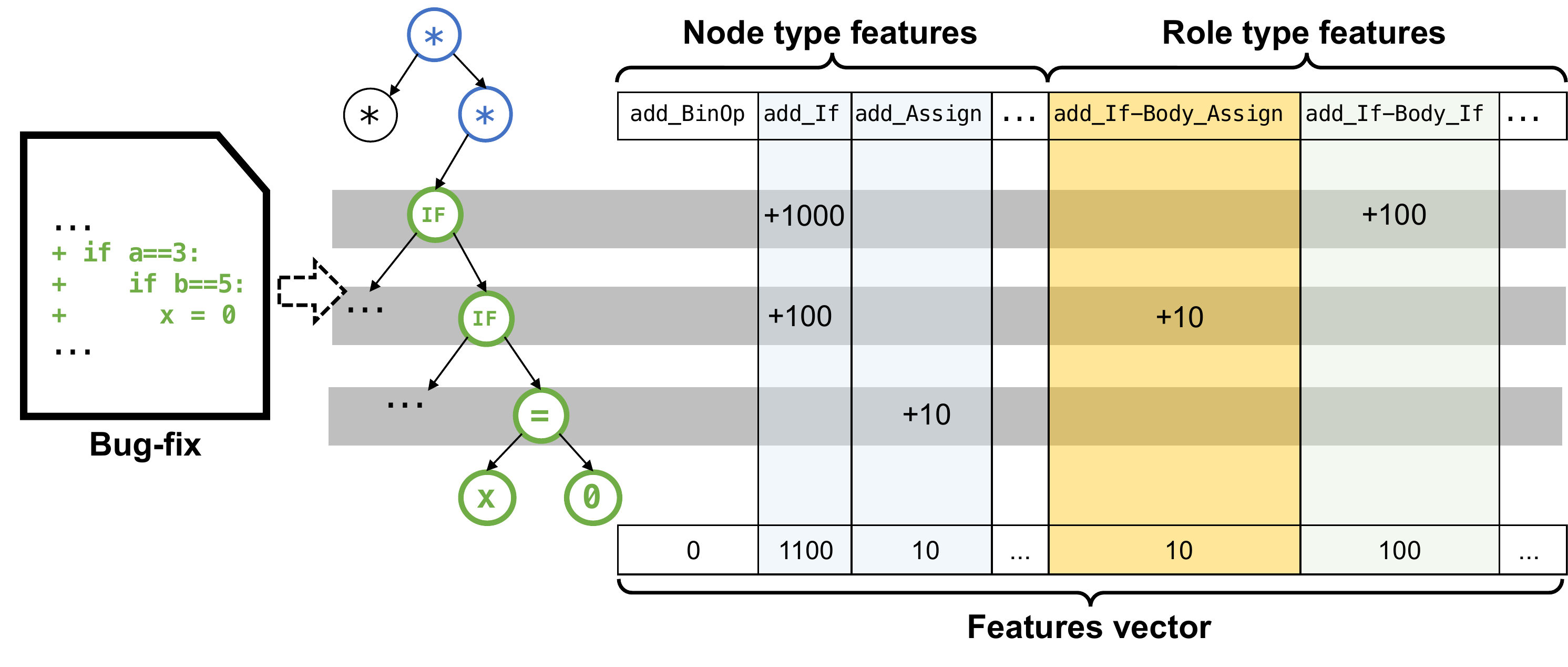}
  \caption{Example of hunk features vector.}
  \label{fig:weighted_feature_example}
\end{figure}

\figurename~\ref{fig:weighted_feature_example} shows an example of a feature vector. The initial \emph{if} node (represented by the type \emph{add\_If}) is increased by a score with the highest weight ($+1000$), and the corresponding role (\emph{add\_If-Body\_If}) is also increased ($+100$). The inner \emph{if} node increases the type feature \emph{add\_If} with a lower weight than the previous \emph{if} node ($+100$) since it is a nested node. In this way, we give greater importance to the fact that the bug-fix is changing the outer \emph{if}, and we give less emphasis to the contents of the \emph{if} (e.g., the content may be another \emph{if}, or other kind of Python statements).
 
We extended \emph{PySA} to automatically compute the features from the hunks. The resulting dataset consists of 5,890 (Nova), 4,261 (Neutron), and 5,930 (Cinder) hunks (dataset rows);
and 948 (Nova), 996 (Neutron), and 1,019 (Cinder) features (dataset columns). We are only considering features for programming constructs that actually occurred in the source code changes of our dataset (as features for unused constructs result in a series of zero values).

We extracted a set of additional features for representing also the code surrounding the bug-fix: \emph{outer} and \textit{inner} \textit{context features}.

\textbf{Outer context features} are extracted from the list of context nodes of the hunk, which includes all the ancestors of the plus and minus nodes (\emph{cfr.} \subsectionautorefname~\ref{subsec:hunk_extraction}).
There are two kinds of outer features:
\begin{itemize}[leftmargin=4mm]

\item \textbf{Features of the including \emph{scoped node}}. These features are related to, and computed from, the hunk's closest ancestor node with {\lmttfont{FunctionDef}}, {\lmttfont{ClassDef}}, or {\lmttfont{Module}} as type. The module, class, or function definition opens a new local scope in the language definition.
These context features describe the including \emph{scoped node} in terms of size (\emph{i.e.,}, number of children).  
In total, there are 6 numeric features defined as { \lmttfont{ctx\_<block\_type>\_size}}, where \texttt{block\_type} can be any of \emph{ClassDef}, \emph{FunctionDef\_args}, \emph{FunctionDef\_body}, or \emph{Module}, plus the boolean feature {\lmttfont ctx\_FunctionDef\_private} (a feature that indicates whether the function is intended for private use only, \emph{i.e.,} its name begins with an underscore).

\item \textbf{Features of the closest ancestor}. These features reflect the type of AST node that is closest to the hunk. The possible ancestors are nodes for iteration ({\lmttfont{For}} and {\lmttfont{While}}), selection ({\lmttfont{If}}), assignment statements ({\lmttfont{Assign}}), definitions ({\lmttfont{ClassDef}}, {\lmttfont{FunctionDef}}, and {\lmttfont{Module}}), exception handling nodes ({\lmttfont{TryExcept}} and {\lmttfont{TryFinally}}), and other kinds of expression statements, including {\lmttfont{Attribute}}, {\lmttfont{BinOp}}, {\lmttfont{BoolOp}}, {\lmttfont{Call}}, {\lmttfont{Return}}, {\lmttfont{Subscript}}. 
There are 15 boolean features defined as {\lmttfont{ctx\_including\_<node\_type>}}, which are all 0s except for the type of the statement that includes the bug-fix.
Furthermore, we have a numeric feature, defined as {\lmttfont{ctx\_including\_node\_size}}, which indicates the number of children of the ancestor node.

\end{itemize}


\textbf{Inner context features} are extracted from the AST nodes below the hunk, in order to provide information on the types of elements (function calls, assignments, arithmetic operations, etc.) that appear in bug-fixed code  (e.g., the block of statements that is surrounded by a new \emph{if}). 
In total, there are 370 features for the inner context, defined as 
{ \lmttfont{ctx\_inner\_<add|rem>\_<node\_type>\_count}}, and
{ \lmttfont{ctx\_inner\_<add|rem>\_<role>\_count}}, 
where \texttt{node\_type} is one of the 89 AST node types in the Python language grammar, and \texttt{role} is one of the 98 AST roles in the grammar.
In total, we computed 232 context features across Nova, Neutron, and Cinder.

\subsection{Hunks Clusterization}
\label{subsec:hunk_clusterization}

In this section, we describe all the choices made for categorizing the hunks found after the  \textit{Hunks Extraction} and \textit{Feature Extraction} phases. Our main objective is to find categories that reflect what has been changed by bug-fixes. We adopt clustering to discover categories with respect to the programming constructs and entities that appear in the hunks (represented by the features discussed in the previous subsection). 

In particular, we applied a hierarchical clustering algorithm, in order to scale to such large datasets, which consists of thousands of samples. We configured the clustering algorithm to use the Euclidean distance and single linkage. To validate the quality of this configuration, we computed the \textbf{cophenetic correlation coefficient} \cite{sokal1962comparison}, which is a measure of how faithfully a dendrogram preserves the pairwise distances between the original unmodeled data point. Hierarchical clustering is an iterative process, in which the closest pair of clusters are merged into one cluster, which replaces the previous pair. Then, the distance matrix is updated by removing the rows/columns of the deleted pair and adding a new row/column for the merged cluster. In some degenerate cases, the new distances in the new row/column may not faithfully be representative of the distances of the previous pair of clusters. The cophenetic coefficient computes the correlation between the new and the previous distances in order to detect such cases. The closer to 1 is the cophenetic coefficient, the more the clustering algorithm preserves correctly the distances between clusters.

The resulting cophenetic coefficients for the datasets are 0.87 (Nova), 0.86 (Neutron), and 0.9 (Cinder), which are leading to consider the configuration good enough for obtaining accurate clustering. 

To determine the natural divisions of the dataset into clusters, we further analyzed the \textbf{inconsistency coefficient} \cite{jain1988algorithms} of the dendrogram links. 
These coefficients compare the height of the link with the average height of other links at the same level of the hierarchy. 
A large coefficient denotes that two ``diverse'' clusters were forcefully merged by the hierarchical clustering algorithms. 
Thus, the links with a higher \emph{inconsistency coefficient} are good candidates for identifying a division of the data into clusters.
Clusters are formed when a node and all of its sub-nodes have an inconsistency value less than a cut-off threshold $c$. 
All leaves at or below the node are grouped into the same cluster.

To identify the cut-off threshold $c$, we first analyze the distribution of the inconsistency coefficients across all links in the dendrogram. 
We obtain such distribution by using the automatic binning algorithm provided by Matlab \cite{matlabhistogram}. 
The binning algorithm divides the distribution among bins of fixed size. 
The algorithm chooses the bin width by adopting a mix of heuristics and well-known algorithms, such as Scott and Freedman-Diaconis rules. 
Our aim is to have a clusterization such that the clusters are not too specialized but they catch the coarse-grained pattern in the code change.
The binning algorithm provides us with intervals (bins) that discretize the values of inconsistency, giving us a hint on how many nodes we preserve if cutting at a certain inconsistency. 
Then, since we want to consider only the greatest differences, we chose $c$ as the lower edge of the last bin.
Thus, only the highest inconsistency values are preserved, aggregating the other nodes of the dendrogram in large clusters.
In our dataset, we obtain $c = 1.15$ both for Nova and Cinder, and $c = 1.1$ for Neutron. 
Since we want to focus on recurring bug patterns, we only consider the largest clusters, by considering the ones with a number of elements greater than a threshold on the cluster distribution (respectively 15, 10, and 15 elements for Nova, Neutron, and Cinder). We obtained $46$ clusters for Nova, $22$ clusters for Neutron, and $43$ clusters for Cinder.

Finally, every cluster has been manually analyzed by two authors (or more, in the case of disagreement) to assess whether the cluster actually represents a bug-fix. 
For every cluster, we manually inspect a sample of $n$ changes in the clusters (in our empirical analysis, we consider $n=5$), and divide the clusters into three categories:

\begin{itemize}[leftmargin=4mm]
    \item \textbf{BUG-FIX} changes, which represent fixes to bugs. We classify a cluster for that category if a majority of changes in the sample actually fixes the behavior of the software, according to the description of the bug and to the nature of the change.

    \item \textbf{FIX-INDUCED} changes, which represent code changes that are required to support a bug-fixing change, but do not represent themselves the actual bug-fix. For example, if the bug-fixing code uses a new input parameter to a method, the signature of the method and the call sites to the method must be also changed as a consequence of the bug-fix.
    
    \item \textbf{REFACTORING}, in which code changes were made for purposes that do not modify the behavior of the software (e.g., better readability or modularization).

\end{itemize}

Finally, once the manual analysis confirms that a cluster represents a bug-fixing change, we attribute a label and a brief description of the cluster, and we consider the cluster for the next analysis of the context.

\subsection{Context Features Analysis}
\label{subsec:context_feature_analysis}

The objective of this analysis is to investigate the hypothesis that bug-fixing changes tend to occur in specific code contexts. In particular, we want to study what are the context features that are representative of a bug-fix change pattern in order to answer the following research question: \emph{Is the context relevant in the characterization of bug-fix changes?}

To answer this question, we compare the mean of each context feature within a bug-fix cluster with the mean of a control group, represented by the whole dataset (including both changes due to bugs, and other changes). To achieve this, we tested the null hypothesis that there is no difference between the bug fix pattern group and the control group, by means of the Dunn’s statistical test \cite{dunn1964multiple}. We used the Dunn’s test as it is robust with respect to groups of uneven size and it is a non-parametric test \cite{dunntest}. Then, we selected all the context features that have a mean statistically different from the control group with a confidence level of 95\%. If, as a result of this process, we find that there is at least one relevant context feature for every cluster, then we can answer affirmatively to the research question.

Moreover, we quantitatively analyzed the context features which resulted relevant by means of summary indicators, i.e., the average, the coefficient of variation (which is defined as the ratio of the standard deviation to the mean), and the distribution quantiles, in order to gain insights on the context conditions that are common to the majority of the bugs included in each bug pattern.

\section{Analysis of \emph{what} is changed by a bug-fix}
\label{sec:what_changes}

\begin{figure}
    \centering
    \includegraphics[width=.7\linewidth]{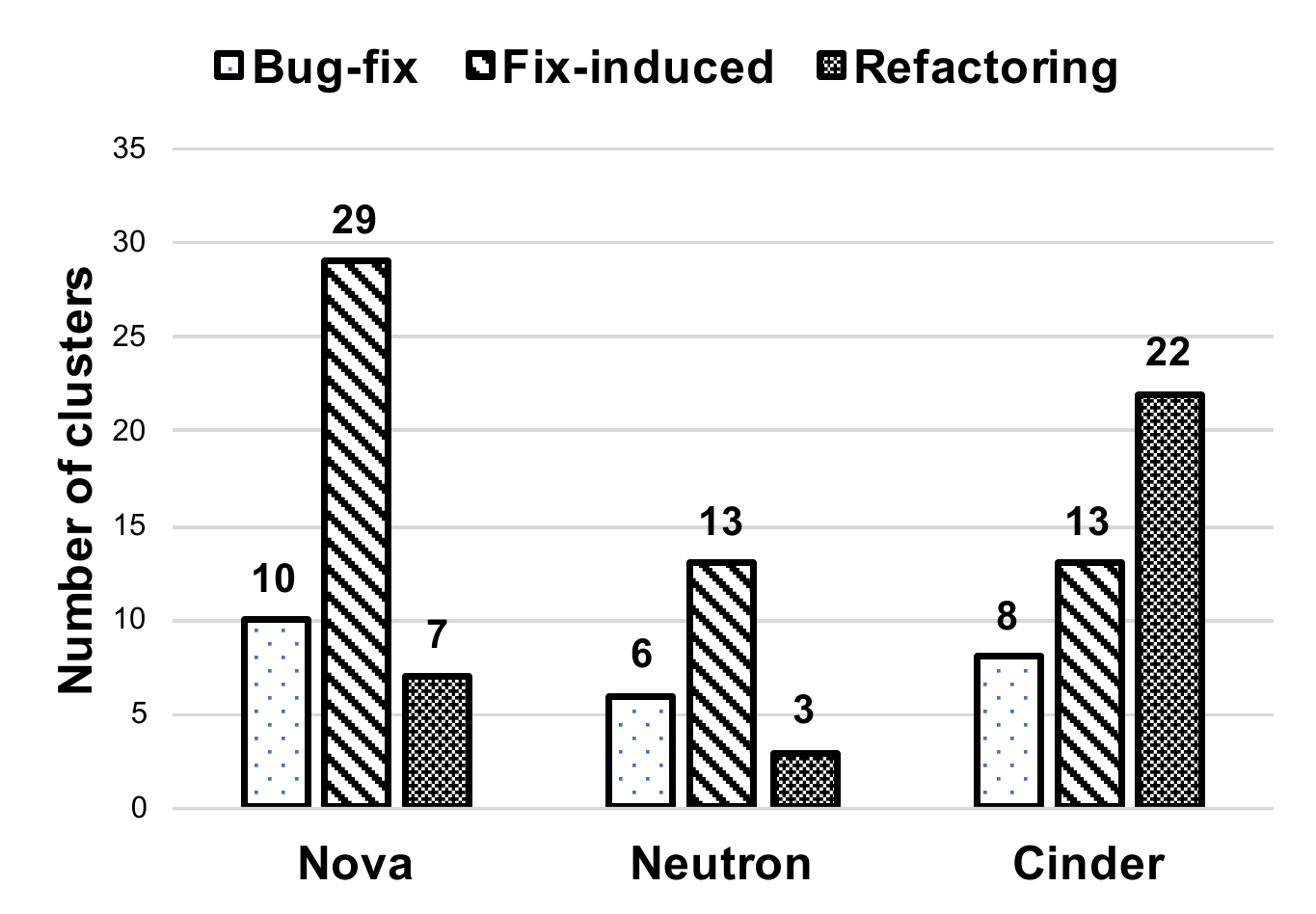}
    \caption{Types of clusters found in Nova, Neutron, and Cinder.}
    \label{fig:clusters_summary}
\end{figure}

In this section, we analyze the bug-fixing patterns obtained by means of clustering. 
We first consider the classification of the clusters between bug-fixes and non-bug-fixes (i.e., FIX-INDUCED and REFACTORING). 
\figurename~\ref{fig:clusters_summary} shows the distribution of the clusters found for the Nova, Neutron, and Cinder projects. The portion of BUG-FIX clusters across the three projects is similar. We found a high number of patterns that were either FIX-INDUCED or REFACTORING changes, with differences across the projects. These changes were included in the same commits for bug-fixes, in which developers took code reviews of bug-fixes as opportunities for also improving the internal quality (e.g., readability and maintainability) of the source code. Thus, both the bug-fixes and the refactoring changes end up in the same commit and get merged in the same revision. 
Since these non-bug-fixes patterns come in a high number, we needed to identify and remove them from our analysis in order to focus on bugs. 
Therefore, we caution other researchers interested in bug-fixing changes to carefully discriminate between bug- and refactoring- related changes, in order to provide more meaningful results for software testing and repair purposes. 
%







%

%

\begin{figure}
  \centering
  \includegraphics[width=.7\linewidth]{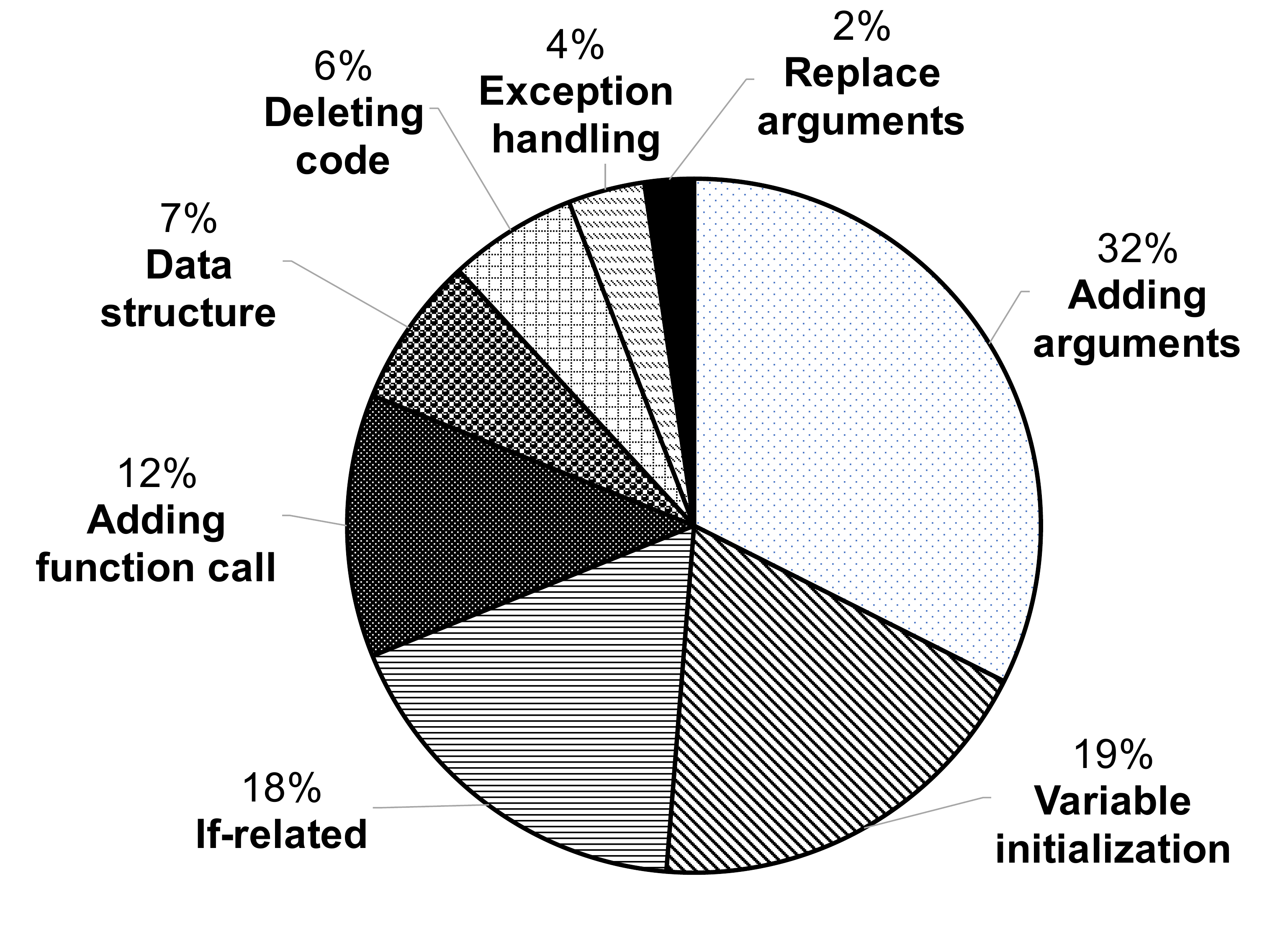}
  \caption{BUG-FIX categories distribution.}
  \label{fig:clusters_bug-fix_pie}
\end{figure}

We focus our analysis on better understanding the BUG-FIX clusters. \tablename~\ref{tab:openstack_clusters} provides the detailed list of clusters, along with a brief description. We also present (\figurename~\ref{fig:clusters_bug-fix_pie}) the BUG-FIX clusters by dividing them into 8 categories, according to the syntactic changes introduced by the bug-fix. Almost half of the bug clusters are related to function calls (e.g., adding new function calls, or new arguments to a function call); the other clusters involve changes to the control flow, data structure initialization, exception handling, etc..

In the following, we first describe more in detail these categories with representative examples of recurrent patterns. Then, we summarize the main findings at the end of this section.

\begin{table}[ht]
    
    \centering
    
    \caption{Bug-fix patterns clusters in Nova, Neutron, and Cinder.}
    \label{tab:openstack_clusters}
    \scalebox{0.85}{
    \begin{tabular}{c c c p{42mm}}
    
    \toprule
    \textbf{Cluster ID} & \textbf{Size} & \textbf{Category} & \textbf{Description}                                        \\ \toprule \\
    
    \rowcolor{LightGrey} &  & \textbf{Nova} &   \\ \\

    {\lmttfont nova\_228}  & 148 & Variable initialization & Add variable initialized to a constant value  \\
    {\lmttfont nova\_131}  & 61  & Adding arguments & Add one variable as keyword parameter to function call                          \\
    {\lmttfont nova\_128}  & 45  & Adding arguments & Add object attribute as keyword parameter to function call                        \\
    {\lmttfont nova\_132}  & 26  & Adding function call & Remove variable as keyword parameter from function call                                        \\
    {\lmttfont nova\_474}  & 22 & If-related & Surround expression with If                                                \\
    {\lmttfont nova\_1099} & 21  & Adding function call & Remove object attribute variable            \\
    {\lmttfont nova\_1097} & 19  & Adding arguments & Add one variable as keywords parameters to multiple function calls\\
    {\lmttfont nova\_597}  & 16  & If-related & Surround instructions with If                                           \\
    {\lmttfont nova\_629}  & 16  & If-related & Add boolean operator in condition                                          \\
    {\lmttfont nova\_197}  & 15  & If-related & Add If and its body                                                        \\ \\
    
    \rowcolor{LightGrey} &  & \textbf{Neutron} &   \\ \\

    {\lmttfont neutron\_119} & 61  & Adding arguments & Add boolean as keyword parameter in function call \\
    {\lmttfont neutron\_13}  & 45  & Adding function call & Add function call with 1 parameter \\
    {\lmttfont neutron\_14}  & 17  & Adding function call & Add function call with no parameters \\
    {\lmttfont neutron\_132} & 15 & Data structure & Add new (key, value) to a dictionary\\
    {\lmttfont neutron\_23}  & 14  & Adding function call & Add function call with 2 parameters \\
    {\lmttfont neutron\_20}  & 10  & If-related & Add If with return statement in body \\ \\
    
    \rowcolor{LightGrey} &  & \textbf{Cinder} &   \\ \\

    {\lmttfont cinder\_115}   & 48  & Adding arguments & Add variable as keyword parameter to function call \\
    {\lmttfont cinder\_621}   & 40  & Data structure & Add new (key, value) to a dictionary \\
    {\lmttfont cinder\_438}   & 35  & If-related & Add assign and add If with its body \\
    {\lmttfont cinder\_583}   & 28 & Exception handling & Surround function call with Try-Except block \\
    {\lmttfont cinder\_627}   & 24  & If-related & Replace boolean expression with function call in If condition \\
    {\lmttfont cinder\_14}    & 20  & Adding function call & Add function call \\
    {\lmttfont cinder\_542}   & 20  & Move function call & Move function call with in a new position  \\
    {\lmttfont cinder\_1168}  & 18  & Replace arguments & Modify constant string parameter in function call \\ 
    
    \bottomrule

\end{tabular}
}
\end{table}

\noindent
\textbf{Adding arguments to function calls}. Several clusters from Nova (e.g., \textbf{{\lmttfont nova\_131}}, \textbf{{\lmttfont nova\_128}}, \textbf{{\lmttfont nova\_1097}}), Neutron (e.g., \textbf{{\lmttfont neutron\_119}}), and Cinder (e.g., \textbf{{\lmttfont cinder\_115}}) are related to fixes that add a new parameter of a function call. In these cases, the developers accidentally forgot to add a variable or an expression as parameter of a function call. 
An example of this kind of bug-fix is showed in Listing \ref{lst:add_param_var_label}, in which the developer adds a variable as input parameter:

\begin{minipage}{.45\textwidth}

\begin{lstlisting}[language=diff, caption={Add simple variable as parameter of a function call. Change No.: 468269. URL: \url{https://review.openstack.org/c/468269/6/nova/virt/libvirt/driver.py}, line 7174}, label={lst:add_param_var_label}, captionpos=b]
- instance_domains = self._host.list_instance_domains()
+ instance_domains = self._host.list_instance_domains(only_running=False)
\end{lstlisting}

\end{minipage}

Another example (Listing \ref{lst:add_param_var_attribute_label}) shows a bug-fix in which the developer added a more elaborated expression (an \emph{attribute} of an object) as parameter of a function call.

\begin{minipage}{.45\textwidth}

\begin{lstlisting}[language=diff,caption={Add an attribute of an object as parameter of a function call. Change No.: 526823. URL: \url{https://review.openstack.org/c/526823/18/nova/scheduler/client/report.py}, line 1624}, label={lst:add_param_var_attribute_label}, captionpos=b]
- r = self.post('/allocations', payload, version=POST_ALLOCATIONS_API_VERSION)
+ r = self.post('/allocations', payload, version=POST_ALLOCATIONS_API_VERSION, global_request_id=context.global_id)       
\end{lstlisting}

\end{minipage}

These examples emphasize that the omissions occurred in functions with optional parameters (such as, optional objects representing a ``context'' for the method and for the resource), and with boolean flags for enabling special behaviors in the function. This relaxed parameter passing is syntactically valid in the Python language, and is extensively used in all OpenStack projects.

\noindent
\textbf{Variable initialization}. The highest number of recurring bug-fixes belong to the cluster \textbf{{\lmttfont nova\_228}}, which includes fixes that add the initialization of a variable, e.g., using a boolean, a null object, or a constant string. 
For example, Listing \ref{lst:add_constant_attribute} shows that the developer forgot to add the attribute \textit{RUN\_ON\_REBUILT} to the class {\lmttfont DiskFilter}. 
In this case, the bug description points out that the change fixed an issue that occurred when a new image was provided and the instance had to be rebuilt, but Nova omitted to validate the existing pool of hosts excluding them from being scheduled.

\begin{minipage}{.45\textwidth}

\begin{lstlisting}[language=diff, caption={Add global variable to the class definition. Change No.: 523212. URL: \url{https://review.openstack.org/c/523212/2/nova/scheduler/filters/disk\_filter.py}, line 31}, label={lst:add_constant_attribute}, captionpos=b]
 class DiskFilter(filters.BaseHostFilter):
     """Disk Filter with over subscription flag."""
+    RUN_ON_REBUILD = False
\end{lstlisting}

\end{minipage}

In general, variable initialization has been a recurring bug pattern in previous studies on C and Java \cite{duraes2006emulation}. In our analysis, we found that these issues were recurring specifically for the Neutron project, where developers often adopted global and class-level variables for controlling the configuration of the Neutron server.


\noindent
\textbf{If-related fixes}. These changes fix the code by modify the control flow, such as: by surrounding an existing statement, or block of statements, with an \emph{if} construct; by adding a new statement or block of statements together with an \emph{if} construct; and by  adding a new boolean condition to an existing one. We found clusters of this kind of changes among Nova (\textbf{{\lmttfont nova\_474}}, \textbf{{\lmttfont nova\_597}}, \textbf{{\lmttfont nova\_629}}, and \textbf{{\lmttfont nova\_197}}), Neutron (\textbf{{\lmttfont neutron\_20}}), and Cinder (\textbf{{\lmttfont cinder\_438}}, \textbf{{\lmttfont cinder\_627}}). These bug-fixes handle corner cases in the user inputs and configuration, such as in the examples of Listing~\ref{lst:if_single_stmt_label} and Listing~\ref{lst:if_add_condition_label}.




\begin{minipage}{.45\textwidth}
\begin{lstlisting}[language=diff, caption={Surround single statement with if construct. Change No.: 442736. URL: \url{https://review.openstack.org/c/442736/27/nova/compute/manager.py}, line 1307}, label={lst:if_single_stmt_label}, captionpos=b]
+ if not CONF.workarounds.disable_group_policy_check_upcall: 
    _do_validation(context, instance, group_hint)    
\end{lstlisting}
\end{minipage}



\begin{minipage}{.45\textwidth}

\begin{lstlisting}[language=diff, caption={Add new condition to an existing one. Change No. 543595. URL: \url{https://review.openstack.org/c/543595/1/nova/scheduler/filters/isolated_hosts_filter.py}, line 64}, label={lst:if_add_condition_label}, captionpos=b]
- if spec_obj.image else None
+ if spec_obj.image and 'id' in spec_obj.image else None
\end{lstlisting}

\end{minipage}

\noindent
\textbf{Adding function calls}. 
The clusters \textbf{{\lmttfont neutron\_13}}, \textbf{{\lmttfont neutron\_14}}, \textbf{{\lmttfont neutron\_23}}, and \textbf{{\lmttfont cinder\_14}} include fixes that add function calls. In these bugs, developers missed a function call, which caused omissions in the workflow of resource management. These issues mostly affected Neutron and Cinder, due to the nature of APIs in these projects. These projects have APIs for propagating across nodes a global view of the state of the data center (such as, the topology of virtual networks), which should be called whereas the state of resources is updated (such as, a network node is added or removed). However, these API calls can be easily omitted since they do not return data that are used afterward; for example, in Listing~\ref{lst:neutron_function_call_1}, a function call was missing after the update of a subnet.






\begin{minipage}{.45\textwidth}

\begin{lstlisting}[language=diff, caption={Add a function call. Change No.: 491409. URL: \url{https://review.openstack.org/c/491409/11/neutron/agent/linux/interface.py}, line 125}, label={lst:neutron_function_call_1}, captionpos=b]
if cidr == default_ipv6_lla: 
+   cidrs.discard(cidr)
    continue
\end{lstlisting}

\end{minipage}


%


\noindent
\textbf{Data structure-related fixes}. The clusters \textbf{{\lmttfont neutron\_132}} and \textbf{{\lmttfont cinder\_621}} include fixes that add a new pair {\lmttfont(key,value)} to a Python dictionary (i.e., the Python data type for associative arrays), in order to fix the layout of the data structure. The OpenStack projects make extensive use of complex data structures to represent the several attributes of virtual resources, such as instances, images, and so on. For example, Listing~\ref{lst:neutron_dict_issue} shows a bug-fix that adds a new entry in a dictionary that represents an \textit{ARP} table in a Neutron component.

\begin{minipage}{.45\textwidth}

\begin{lstlisting}[language=diff, caption={Add new key value to a dictionary. Change No.: 554729. URL: \url{https://review.openstack.org/c/554729/3/neutron/db/l3\_dvr\_db.py}, line 918}, , label={lst:neutron_dict_issue}, captionpos=b]
    arp_table = {'ip_address': ip_address,
                        'mac_address': mac_address,
                        'subnet_id': subnet,
+                       'nud_state': nud_state}
\end{lstlisting}

\end{minipage}

\noindent
\textbf{Exception handling fixes}. The bug-fixes in the cluster \textbf{{\lmttfont cinder\_583}} address missing exceptions, by adding try-except blocks around existing code. Listing~\ref{lst:try_except_fix} shows an example of bug-fix that addresses the case of an exception raised by {\lmttfont NetApp} (one of the several backend drivers supported by Cinder) when the callers try to delete a volume that does not exist.

\begin{minipage}{.45\textwidth}

\begin{lstlisting}[language=diff, caption={Surround function call with Try-Except block. Change No.: 491962. URL: \url{https://review.openstack.org/c/491962/1/cinder/volume/drivers/netapp/dataontap/block\_base.py}, line 284}, , label={lst:try_except_fix}, captionpos=b]
+   try:
        self.zapi_client.destroy_lun(metadata['Path'])
+   except netapp_api.NaApiError as e:
+       if e.code == netapp_api.EOBJECTNOTFOUND:
+           LOG.warning(_LW("Failure deleting LUN %(name)s." " %(message)s"), {'name': lun_name, 'message': e})
+       else:
+           error_message = (_('A NetApp Api Error occurred: %s') % e)
+           raise exception.NetAppDriverException(error_message)
\end{lstlisting}

\end{minipage}


\noindent
\textbf{Replace string arguments in function call}. The cluster \textbf{{\lmttfont cinder\_1168}} includes bug-fixes that modify a string argument in a function call.  Listing~\ref{lst:replace_string_arg_fix} shows an example in which the string parameter is used to represent a path in the filesystem. 
These issues were recurrent due to the frequent use of external Linux commands in OpenStack. For example, Cinder uses administration utilities for handling storage volumes (e.g., {\lmttfont tgtadm} for SCSI, {\lmttfont ietadm} for iSCSI, etc.), and basic Linux commands for handling files (e.g., {\lmttfont touch}, {\lmttfont tee}, etc.).

\begin{minipage}{.45\textwidth}

\begin{lstlisting}[language=diff, caption={Incorrect string parameter. Change No.: 491962. URL: \url{https://review.opendev.org/c/465961/2/cinder/volume/drivers/ibm/gpfs.py}, line 211}, , label={lst:replace_string_arg_fix}, captionpos=b]
-   (out, err) = self.gpfs_execute('mmlsconfig', 'clusterId', '-Y')
+   (out, err) = self.gpfs_execute(self.GPFS_PATH + 'mmlsconfig', 'clusterId', '-Y')
\end{lstlisting}

\end{minipage}


\vspace{2pt}
\noindent
\textbf{Deleting code bug-fixes}. The clusters \textbf{{\lmttfont nova\_132}} and \textbf{{\lmttfont nova\_1099}} include bug-fixes that remove surplus code. For example, in Listing~\ref{lst:remove_param}, the fix removed a parameter in excess ({\lmttfont retry\_on\_request}) from an API call since that argument become deprecated after an update of the class {\lmttfont wrap\_db\_retry} (where retry on request is always enabled). 
In general, surplus code occurred because of regressions, such as, APIs that are deprecated or that adopt new calling conventions, or changes in third-party software that is included in the project, or incorrect new code that is reversed to a previously-working version.



\begin{minipage}{.45\textwidth}

\begin{lstlisting}[language=diff, caption={Remove keyword parameter from function call. Change No. 501073. URL: \url{https://review.openstack.org/c/501073/1/nova/db/sqlalchemy/api.py}, line 64}, label={lst:remove_param}, captionpos=b]
@require_context
-   @oslo_db_api.wrap_db_retry(max_retries=5, retry_on_deadlock=True, retry_on_request=True)
+   @oslo_db_api.wrap_db_retry(max_retries=5, retry_on_deadlock=True)
\end{lstlisting}

\end{minipage}

From our analysis of the clusters, we make the following observations on the general trends that we observed in bug-fixing changes.


\begin{tcolorbox}


\textit{\textbf{Observation 1}}. 
\textbf{A minority of bug-fix patterns matches the ones found in previous studies} on Java and C software.
\end{tcolorbox}
    In particular, this group of bug-fix patterns includes the ones in the \emph{If-related} category, which is one of the larges category found in our analysis (18\% in \figurename~\ref{fig:clusters_bug-fix_pie}). These patterns are consistent with other studies on bug analysis \cite{soto2016deeper, lin2016empirical, osman2014mining, pan2009toward, martinez2013automatically, fluri2008discovering, duraes2006emulation, basso2009investigation}, which found recurring issues that were fixed in the control flow (e.g.,the \emph{checking} and \emph{algorithm} categories in the ODC classification). These patterns were also consistently found across all of the Python projects.
    There were other bug-fix patterns that were similar to the ones found in previous studies, which include bug-fix that \emph{added a function call} and the \emph{replace arguments in a function call} (e.g., the \emph{interface} category in the ODC classification). However, these patterns were not consistent across the projects, as they were only found in Neutron and Cinder, as discussed in the \emph{Observation 3}. 

\begin{tcolorbox}
\textit{\textbf{Observation 2}}. 
\textbf{Several patterns are related to the data structures and rules of the Python language}, and are common across projects.
\end{tcolorbox}

    There were bug-fix patterns dependent on the Python language used for the projects. In particular, the bug-fixes involving \emph{data structures} affected Python {\lmttfont dict}s, which are a common way to represent data in this language. Therefore, it is reasonable to expect that a noticeable share of bug-fixes affect these constructs, such as in their layout or in the contents of the data structures. 
    Moreover, the \emph{adding arguments} bug-fix patterns (32\% in \figurename~\ref{fig:clusters_bug-fix_pie}) seem also favored by the rules of the Python language. Differing from C and Java, the Python language makes easier for developer to have optional parameters in function calls, where a default value is assumed by the function if no parameter is passed at the call site. Therefore, while previous studies \cite{duraes2006emulation, basso2009investigation} found bugs in C and Java software where \emph{wrong} parameters were passed (e.g., using a wrong variable or an incomplete arithmetic expression), in our analysis the parameters were mostly \emph{missing}. These bugs were recurrent across the OpenStack projects. 
 


\begin{tcolorbox}
\textit{\textbf{Observation 3}}. 
\textbf{Most bug-fix patterns are project-specific}, as they are induced by API conventions, by the QA and testing process of the project, and the programming idioms used by developers.
\end{tcolorbox}


    We found several bug-fix patterns that were specific to only some of the OpenStack projects. One of the causes was the design of APIs in the Neutron and Cinder projects, which developers had to call throughout many different places of the codebase in order to keep updated the global state view; these calls were often omitted, and were later added by bug-fixing changes. In the case of the Cinder project, bugs with wrong string parameters were related to invocations of external Linux commands, as these utility programs are often used for system administration purposes. 
    Another cause of project-specific patterns were regressions in Nova, which were fixed by deleting surplus code, going back to a working version of the code. The occurrence of regressions can be related to the testing and QA process of the projects since a less effective process can lead to more regressions that are later addressed by bug-fixing changes. 
    Finally, project-specific patterns were related to the programming idioms adopted in the project. For example, in the Nova project, many bug-fixes initialize a variable with a constant since such variables are often used in the projects for global or class-level configurations.




These observations provide information on which bug-fixing patterns apply to Python software. While some of these patterns are consistent across programming languages and projects, other ones are either influenced by the language or by the nature of the project. When pursuing tasks based on code mutations, such as automated program repair or fault injection, this finding motivates the calibration of code change patterns according to the specific project at hand. In the following section, we focus on the context in which these code changes should be performed.
\section{Analysis of \emph{where} bug-fix changes are made}
\label{sec:where_changes}

In this analysis, we consider the features about the code surrounding the bug-fixing change. Since we have a large number of context features, we simplify the presentation of results, by grouping context features in 17 categories with descriptive names. The first two sets of categories constitute the \textit{outer context}, while the last set of category forms the \textit{inner context}. In particular, we considered:
\begin{itemize}[leftmargin=4mm]

    \item 3 categories to group the feature of the including scoped node: \emph{Class Size}, \emph{Function Size}, \emph{Module Size} (\emph{cfr.} \subsectionautorefname~\ref{subsec:code_change_feature_extraction});

    \item 8 categories to group the features representing the closest ancestor that includes the bug-fix: \emph{Closest Definition}, \emph{Closest Exception}, \emph{Closest Iteration}, \emph{Closest Selection}, \textit{Closest Access}, \textit{Closest Call},  \emph{Closest Assign}, and \emph{Closest Size} (\subsectionautorefname~\ref{subsec:code_change_feature_extraction});

    \item 6 categories to group the inner context features: \emph{Assign Operators}, \emph{Control Flow}, \emph{Data Containers}, \emph{Functions}, \emph{Globals} and \emph{Special Operators} (i.e., Python operators such as \lmttfont{print}, \lmttfont{raise}, \lmttfont{return}, \lmttfont{with}, etc.).
    

\end{itemize}


\tablename~\ref{tab:ctx_nova_summary}, \tablename~\ref{tab:ctx_neutron_summary}, and \tablename~\ref{tab:ctx_cinder_summary} show an overview on what context exists for a bug-fix cluster. Each table presents the {\lmttfont cluster ID} on the columns (see \sectionautorefname~\ref{sec:what_changes} for detailed descriptions of the clusters), and the 17 context feature categories on the rows. In the table cells, the checkmark symbol (\Checkmark{}) points out that a given context feature is statistically relevant for the cluster according to the Dunn test (i.e., the bug-fixes in the cluster show a significant deviation of the metric compared to the norm of the other changes; see \subsectionautorefname~\ref{subsec:context_feature_analysis}). 

%
Every cluster exhibits several context features with relevant deviations. Consequently, the context in which a bug-fix was made should not be overlooked, as it provides characterizing information for the bug-fix pattern. 
Moreover, all of the feature categories for both the outer context and inner context exhibit a statistically-significant relevance for at least one project. Therefore, the answer to the research question in \sectionautorefname~\ref{subsec:context_feature_analysis} is positive: \emph{the context is relevant in the characterization of bug-fix changes}.

For example, by focusing on the outer context features, we can notice that almost in all clusters for Nova and Cinder, the context of a bug is characterized by the \textit{Function Size}. This feature includes the number of function arguments, the size of the function body, and whether the function definition is marked as private. This implies that testing and repair algorithms should seek for functions with large functions in order to apply these code change patterns. 

\begin{figure}
  \centering
  \includegraphics[width=0.7\columnwidth]{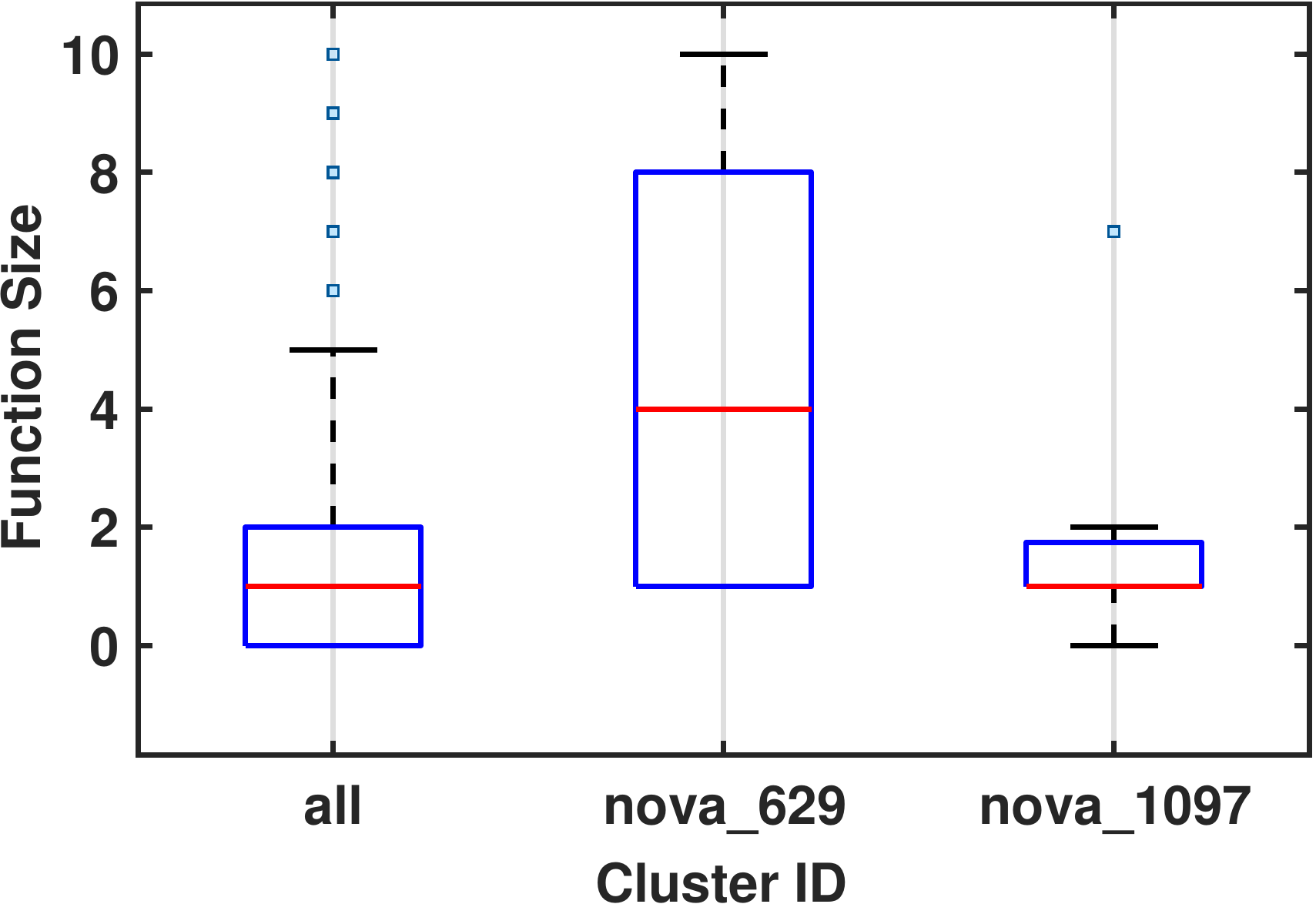}
  \caption{Comparison of the \textit{Function Size} context feature in two clusters against the control group (i.e., the \texttt{all} group).}
  \label{fig:representative_context_feature}
\end{figure}

\figureautorefname{}~\ref{fig:representative_context_feature} shows a visual example for two clusters where the \emph{Function Size} is respectively relevant, and not relevant, for the cluster. The three box plots show the variation of the \textit{Function Size} feature across the whole dataset (first box), the {\lmttfont nova\_629} bug-fix pattern (second box) and the {\lmttfont nova\_1097} bug-fix pattern (third box). There are no relevant differences between the bug-fix pattern {\lmttfont nova\_1097} and the control group (\emph{all}), while the cluster {\lmttfont nova\_629} has a mean which is significantly greater than the control group. Therefore, we consider the \emph{Function Size} feature as statistically relevant to describe the context of the bug fix pattern {\lmttfont nova\_629}, as confirmed in the Table~\ref{tab:ctx_nova_summary} by means of the Dunn test results.

%

Across the OpenStack projects, some context feature classes are more frequent than others. For example, the inner context feature category \textit{Function} describes 6 out of 11 bug patterns in Nova, 6 out of 8 bug patterns in Neutron, and 7 out of 8 bug patterns in Cinder. The metrics in the \textit{Function} category include the number of arguments and keywords were added/removed in a function call, or newly added function calls, and how many function calls appeared in a block inside the bug-fix (e.g., a block of statements that was surrounded by a new \emph{if}). Regardless of the bug-fix pattern (e.g., If-related, Data structure, etc.), in the majority of the cases, the context of bug-fix is characterized by the presence of function calls. 
Another important category of inner context features is \textit{Data Containers}. These features keep track of the presence of special Python data structures (e.g., {\lmttfont dict}s, {\lmttfont list}s, {\lmttfont tuple}s) within the statements that were changed by a bug-fix. For example, the context feature for {\lmttfont dict}s counts how many keys and values appeared in expressions that were added or removed by the bug-fix. Therefore, these bug-fixing patterns tend to occur in the context of complex expressions and data-structure layouts. 

\begin{table*}

\caption{Context features class and BUG-FIX clusters for Nova, Cinder and Neutron.}
\label{tab:ctx_summary}

\begin{subtable}{.40\linewidth}


\caption{Nova}
\label{tab:ctx_nova_summary}

\centering{}%

\scalebox{0.55}{
\begin{tabular}{rccccccccccc}
\toprule 
\multirow{2}{*}{\textbf{\small{}Features}} & \multicolumn{11}{c}{\textbf{\small{}Bug Cluster (Nova Project)}}\tabularnewline
\cmidrule{2-12} \cmidrule{3-12} \cmidrule{4-12} \cmidrule{5-12} \cmidrule{6-12} \cmidrule{7-12} \cmidrule{8-12} \cmidrule{9-12} \cmidrule{10-12} \cmidrule{11-12} \cmidrule{12-12} 
 & \textbf{\small{}128} & \textbf{\small{}131} & \textbf{\small{}132} & \textbf{\small{}197} & \textbf{\small{}228} & \textbf{\small{}238} & \textbf{\small{}474} & \textbf{\small{}597} & \textbf{\small{}629} & \textbf{\small{}1097} & \textbf{\small{}1099}\tabularnewline
\midrule
\midrule 

\textbf{\small{}Module Size} & \textbf{\small{}\Checkmark{}} & \textbf{\small{}\Checkmark{}} & \textbf{\small{}\Checkmark{}} & \textbf{\small{}\Checkmark{}} & \textbf{\small{}\Checkmark{}} & \textbf{\small{}\Checkmark{}} & \textbf{\small{}\Checkmark{}} &  &  & \textbf{\small{}\Checkmark{}} & \textbf{\small{}\Checkmark{}}\tabularnewline
\midrule 

\textbf{\small{}Class Size} & \textbf{\small{}\Checkmark{}} & \textbf{\small{}\Checkmark{}} &  &  & \textbf{\small{}\Checkmark{}} & \textbf{\small{}\Checkmark{}} & \textbf{\small{}\Checkmark{}} &  &  &  & \tabularnewline
\midrule 

\textbf{\small{}Function Size} & \textbf{\small{}\Checkmark{}} & \textbf{\small{}\Checkmark{}} &  & \textbf{\small{}\Checkmark{}} & \textbf{\small{}\Checkmark{}} & \textbf{\small{}\Checkmark{}} & \textbf{\small{}\Checkmark{}} & \textbf{\small{}\Checkmark{}} & \textbf{\small{}\Checkmark{}} &  & \textbf{\small{}\Checkmark{}}\tabularnewline
\midrule 
\midrule

\textbf{\small{}Closest Definition} &  &  &  &  & \textbf{\small{}\Checkmark{}} &  &  &  &  &  & \tabularnewline
\midrule 

\textbf{\small{}Closest Exception} &  &  &  & \textbf{\small{}\Checkmark{}} &  &  & \textbf{\small{}\Checkmark{}} &  &  &  & \tabularnewline
\midrule 

\textbf{\small{}Closest Iteration} &  &  &  &  &  &  & \textbf{\small{}\Checkmark{}} & \textbf{\small{}\Checkmark{}} &  &  & \tabularnewline
\midrule 

\textbf{\small{}Closest Selection} &  &  &  &  &  &  & \textbf{\small{}\Checkmark{}} &  & \textbf{\small{}\Checkmark{}} &  & \tabularnewline
\midrule 

\textbf{\small{}Closest Attribute} & \textbf{\small{}\Checkmark{}} & \textbf{\small{}\Checkmark{}} & \textbf{\small{}\Checkmark{}} &  &  & \textbf{\small{}\Checkmark{}} &  &  & \textbf{\small{}\Checkmark{}} & \textbf{\small{}\Checkmark{}} & \tabularnewline
\midrule 

\textbf{\small{}Closest Call} & \textbf{\small{}\Checkmark{}} & \textbf{\small{}\Checkmark{}} & \textbf{\small{}\Checkmark{}} &  &  & \textbf{\small{}\Checkmark{}} &  &  &  & \textbf{\small{}\Checkmark{}} & \tabularnewline
\midrule 

\textbf{\small{}Closest Assign} &  &  &  &  &  &  &  &  & \textbf{\small{}\Checkmark{}} & \textbf{\small{}\Checkmark{}} & \tabularnewline
\midrule

\textbf{\small{}Closest Size} &  &  &  &  &  & \textbf{\small{}\Checkmark{}} &  &  & \textbf{\small{}\Checkmark{}} &  & \textbf{\small{}\Checkmark{}}\tabularnewline
\midrule 
\midrule 

\textbf{\small{}Assign Operators} &  &  &  & \textbf{\small{}\Checkmark{}} & \textbf{\small{}\Checkmark{}} &  & \textbf{\small{}\Checkmark{}} & \textbf{\small{}\Checkmark{}} & \textbf{\small{}\Checkmark{}} &  & \tabularnewline
\midrule 

\textbf{\small{}Control Flow} &  &  &  & \textbf{\small{}\Checkmark{}} &  &  & \textbf{\small{}\Checkmark{}} & \textbf{\small{}\Checkmark{}} &  &  & \tabularnewline
\midrule 

\textbf{\small{}Data Containers} & \textbf{\small{}\Checkmark{}} &  &  & \textbf{\small{}\Checkmark{}} & \textbf{\small{}\Checkmark{}} & \textbf{\small{}\Checkmark{}} & \textbf{\small{}\Checkmark{}} & \textbf{\small{}\Checkmark{}} & \textbf{\small{}\Checkmark{}} & \textbf{\small{}\Checkmark{}} & \textbf{\small{}\Checkmark{}}\tabularnewline
\midrule 

\textbf{\small{}Function} & \textbf{\small{}\Checkmark{}} & \textbf{\small{}\Checkmark{}} & \textbf{\small{}\Checkmark{}} &  &  &  & \textbf{\small{}\Checkmark{}} & \textbf{\small{}\Checkmark{}} &  & \textbf{\small{}\Checkmark{}} & \tabularnewline
\midrule

\textbf{\small{}Globals} &  &  &  &  &  &  &  & \textbf{\small{}\Checkmark{}} &  & \textbf{\small{}\Checkmark{}} & \tabularnewline
\midrule

\textbf{\small{}Special Operators} &  &  &  &  &  &  &  &  &  &  & \tabularnewline
\bottomrule

\end{tabular}
}

\end{subtable}%
\begin{subtable}{.30\linewidth}

\caption{Neutron}
\label{tab:ctx_neutron_summary}

\centering{}%
\scalebox{0.55}{
\begin{tabular}{rcccccccc}
\toprule 
\multirow{2}{*}{\textbf{\small{}Features}} & \multicolumn{8}{c}{\textbf{\small{}Bug Cluster (Neutron Project)}}\tabularnewline
\cmidrule{2-9} \cmidrule{3-9} \cmidrule{4-9} \cmidrule{5-9} \cmidrule{6-9} \cmidrule{7-9} \cmidrule{8-9} \cmidrule{9-9} 
 & {\small{}13} & {\small{}14} & {\small{}20} & {\small{}23} & {\small{}74} & {\small{}119} & {\small{}132} & {\small{}209}\tabularnewline
\midrule
\midrule 

\textbf{\small{}Module Size} &  &  &  & {\small{}\Checkmark{}} &  &  &  & \tabularnewline
\midrule 

\textbf{\small{}Class Size} &  &  & {\small{}\Checkmark{}} &  &  &  &  & \tabularnewline
\midrule 

\textbf{\small{}Function Size} & {\small{}\Checkmark{}} & {\small{}\Checkmark{}} &  &  &  & {\small{}\Checkmark{}} &  & \tabularnewline
\midrule 
\midrule

\textbf{\small{}Closest Definition} &  & {\small{}\Checkmark{}} & {\small{}\Checkmark{}} & {\small{}\Checkmark{}} &  &  &  & {\small{}\Checkmark{}}\tabularnewline
\midrule 

\textbf{\small{}Closest Exception} &  &  &  &  &  &  &  & \tabularnewline
\midrule 

\textbf{\small{}Closest Iteration} & {\small{}\Checkmark{}} & {\small{}\Checkmark{}} &  &  & {\small{}\Checkmark{}} &  &  & \tabularnewline
\midrule 

\textbf{\small{}Closest Selection} & {\small{}\Checkmark{}} &  &  &  &  &  &  & \tabularnewline
\midrule 

\textbf{\small{}Closest Attribute} &  &  &  &  &  & {\small{}\Checkmark{}} &  & \tabularnewline
\midrule 

\textbf{\small{}Closest Call} &  &  &  &  &  & {\small{}\Checkmark{}} &  & \tabularnewline
\midrule 

\textbf{\small{}Closest Assign} &  &  &  &  &  &  & {\small{}\Checkmark{}} & {\small{}\Checkmark{}} \tabularnewline
\midrule

\textbf{\small{}Closest Size} &  &  &  &  &  &  & {\small{}\Checkmark{}} & \tabularnewline
\midrule 
\midrule 

\textbf{\small{}Assign Operators} & {\small{}\Checkmark{}} & {\small{}\Checkmark{}} & {\small{}\Checkmark{}} & {\small{}\Checkmark{}} &  &  &  & \tabularnewline
\midrule

\textbf{\small{}Control Flow} &  &  & {\small{}\Checkmark{}} &  &  &  &  & \tabularnewline
\midrule 

\textbf{\small{}Data Containers} &  &  & {\small{}\Checkmark{}} &  & {\small{}\Checkmark{}} &  &  & {\small{}\Checkmark{}}\tabularnewline
\midrule 

\textbf{\small{}Function} & {\small{}\Checkmark{}} & {\small{}\Checkmark{}} & {\small{}\Checkmark{}} & {\small{}\Checkmark{}} & {\small{}\Checkmark{}} & {\small{}\Checkmark{}} &  & \tabularnewline
\midrule

\textbf{\small{}Globals} &  &  &  &  &  &  & {\small{}\Checkmark{}} & \tabularnewline
\midrule

\textbf{\small{}Special Operators} &  &  &  &  &  &  &  & \tabularnewline
\bottomrule
\end{tabular}
}
\end{subtable}%
\begin{subtable}{.30\linewidth}

\caption{Cinder}
\label{tab:ctx_cinder_summary}

\centering{}%
\scalebox{0.55}{
\begin{tabular}{rcccccccc}
\toprule 
\multirow{2}{*}{\textbf{\small{}Features}} & \multicolumn{8}{c}{\textbf{\small{}Bug Cluster (Cinder Project)}}\tabularnewline
\cmidrule{2-9} \cmidrule{3-9} \cmidrule{4-9} \cmidrule{5-9} \cmidrule{6-9} \cmidrule{7-9} \cmidrule{8-9} \cmidrule{9-9} 
 & {\small{}14} & {\small{}115} & {\small{}438} & {\small{}542} & {\small{}583} & {\small{}621} & {\small{}627} & {\small{}1168}\tabularnewline
\midrule
\midrule 

\textbf{\small{}Module Size} & {\small{}\Checkmark{}} & {\small{}\Checkmark{}} & {\small{}\Checkmark{}} & {\small{}\Checkmark{}} &  &  & {\small{}\Checkmark{}} & {\small{}\Checkmark{}}\tabularnewline
\midrule 

\textbf{\small{}Class Size} & {\small{}\Checkmark{}} &  &  &  & {\small{}\Checkmark{}} & {\small{}\Checkmark{}} & {\small{}\Checkmark{}} & {\small{}\Checkmark{}}\tabularnewline
\midrule 

\textbf{\small{}Function Size} & {\small{}\Checkmark{}} & {\small{}\Checkmark{}} & {\small{}\Checkmark{}} &  & {\small{}\Checkmark{}} & {\small{}\Checkmark{}} & {\small{}\Checkmark{}} & {\small{}\Checkmark{}}\tabularnewline
\midrule 
\midrule

\textbf{\small{}Closest Definition} &  &  & {\small{}\Checkmark{}} & {\small{}\Checkmark{}} & {\small{}\Checkmark{}} &  &  & \tabularnewline
\midrule 

\textbf{\small{}Closest Exception} &  &  &  &  &  &  &  & \tabularnewline
\midrule 

\textbf{\small{}Closest Iteration} &  &  &  &  &  &  &  & \tabularnewline
\midrule 

\textbf{\small{}Closest Selection} & {\small{}\Checkmark{}} &  &  &  & {\small{}\Checkmark{}} &  & {\small{}\Checkmark{}} & \tabularnewline
\midrule 

\textbf{\small{}Closest Attribute} &  & {\small{}\Checkmark{}} &  &  &  & {\small{}\Checkmark{}}  &  & {\small{}\Checkmark{}} \tabularnewline
\midrule 

\textbf{\small{}Closest Call} &  & {\small{}\Checkmark{}} &  &  &  & {\small{}\Checkmark{}}  &  & {\small{}\Checkmark{}} \tabularnewline
\midrule 

\textbf{\small{}Closest Assign} &  &  &  &  &  &   &  &  \tabularnewline
\midrule

\textbf{\small{}Closest Size} &  &  & {\small{}\Checkmark{}} & {\small{}\Checkmark{}} &  & {\small{}\Checkmark{}} &  & \tabularnewline
\midrule 
\midrule 

\textbf{\small{}Assign Operators} & {\small{}\Checkmark{}} &  & {\small{}\Checkmark{}} & {\small{}\Checkmark{}} & {\small{}\Checkmark{}} &  & {\small{}\Checkmark{}} & {\small{}\Checkmark{}}\tabularnewline
\midrule

\textbf{\small{}Control Flow} &  &  & {\small{}\Checkmark{}} & {\small{}\Checkmark{}} & {\small{}\Checkmark{}} &  &  & \tabularnewline
\midrule 

\textbf{\small{}Data Containers} &  &  & {\small{}\Checkmark{}} & {\small{}\Checkmark{}} & {\small{}\Checkmark{}} & {\small{}\Checkmark{}} & {\small{}\Checkmark{}} & \tabularnewline
\midrule 

\textbf{\small{}Function} & {\small{}\Checkmark{}} & {\small{}\Checkmark{}} & {\small{}\Checkmark{}} & {\small{}\Checkmark{}} & {\small{}\Checkmark{}} & {\small{}\Checkmark{}} & {\small{}\Checkmark{}} & \tabularnewline
\midrule

\textbf{\small{}Globals} &  &  & {\small{}\Checkmark{}} &  &  & {\small{}\Checkmark{}} &  & \tabularnewline
\midrule

\textbf{\small{}Special Operators} &  &  &  &  & {\small{}\Checkmark{}} &  &  & \tabularnewline
\bottomrule
\end{tabular}
}

\end{subtable}%

\end{table*}

\begin{figure}
  \centering
  \includegraphics[width=.9\columnwidth]{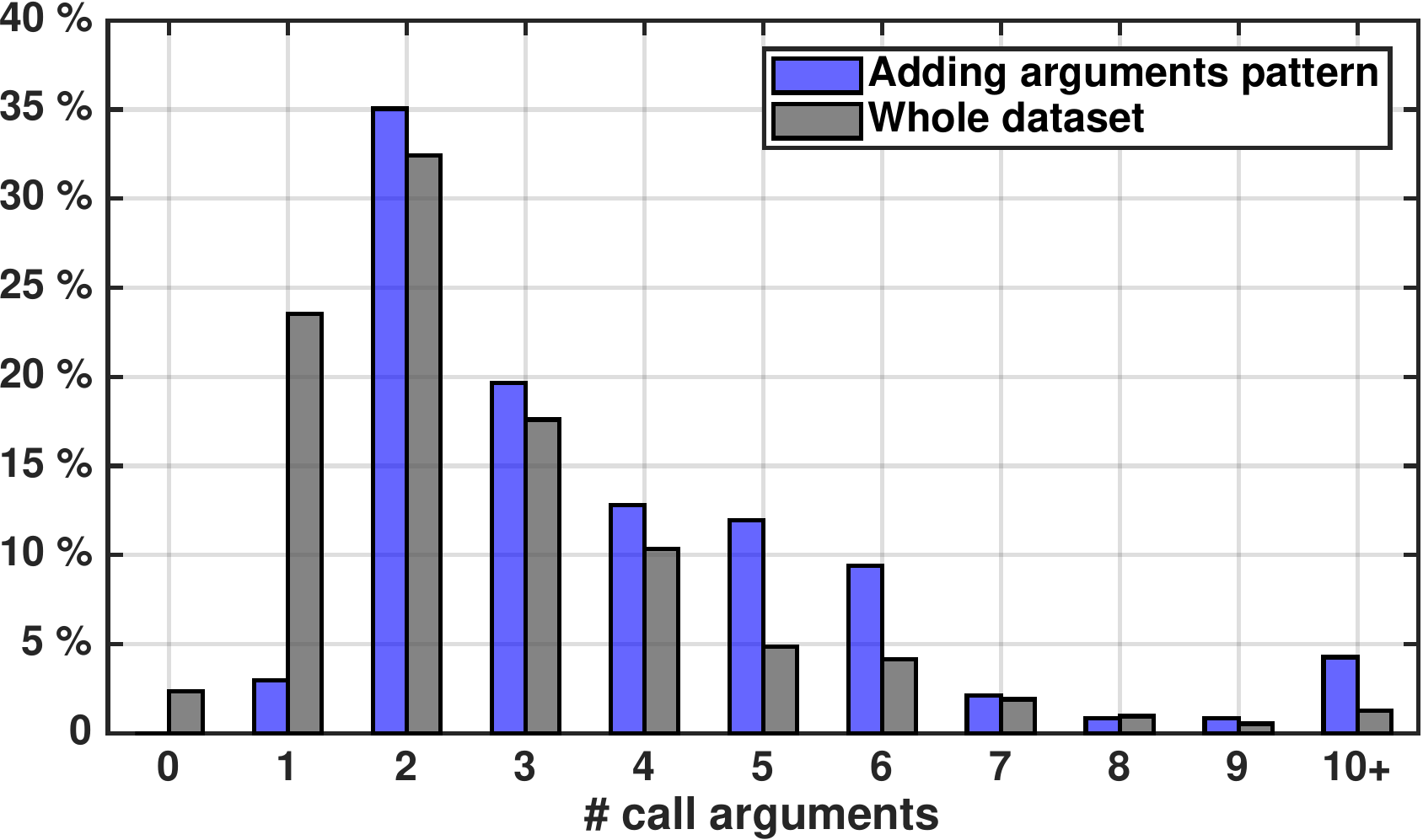}
  \caption{Distribution of the number of arguments for the \textit{Closest call} context feature.}
  \label{fig:including_call_context}
  \vspace{-7mm}
\end{figure}

Similarly, the \textit{Closest Call} and the \textit{Closest Attribute} categories are relevant context features for all of the clusters of the type \textit{Adding arguments} (\tableautorefname{}~\ref{tab:openstack_clusters}), where the bug-fix adds parameters to a function call. These categories bring useful information to characterize the change patterns in these clusters. In particular, the \textit{Closest Call} category includes features for counting the number of arguments in the fixed function call (beyond the argument that is added by the bug-fix). A closer analysis of this feature tells us that the \textit{Adding arguments} bug-fix applies mostly to function calls that already have some parameters. 
\figureautorefname{}~\ref{fig:including_call_context} shows the distribution of the number of arguments of the function call fixed by the changes. On average, the number of arguments settles between 2-3 for the bug-fixing patterns; only in a few cases the fixed function call had 1 or no parameters. This observation improves the characterization of bug-fixing changes that add parameters: it points out that these bug-fixing changes do not uniformly apply to all function calls, and that new parameters tend to be added on function calls that already take several input parameters.





Finally, we can notice differences across the Nova, Neutron, and Cinder projects. For example, if we focus on the features on the scoped nodes that include the change (i.e., module size, class size, and function size), the results show some similarities between Nova and Cinder, and differences between them and the Neutron project. In the Neutron project, it seems that the size of modules, classes and functions, and the number of their arguments, is less relevant than for the Nova and Cinder projects. This result points out that the context for applying code changes needs to be calibrated with respect to the specific project.



\vspace{-2mm}
\section{Threats to validity}
\label{sec:threats_to_validity}
\vspace{-1mm}


We here review the main potential threats that can affect the validity of results, and how we mitigated them.

\textbf{Construct validity} refers to the relationship between the theory and the observation. A threat is that OpenStack revisions include code changes not related to bug-fixes (e.g., new features, documentation, refactoring, etc.). To avoid this threat we selected only revisions having a description with specific keywords (e.g., \emph{fix}, see \sectionautorefname~\ref{sec:methodology}). Since this text is filled out by humans, it is possible to wrongly include in the analysis also non bug-fix changes (e.g., the expression \emph{“fix code programming style”} refers to refactoring changes but includes the term \emph{fix}). To avoid these cases, we classified and excluded these changes by manual inspection during the post-hoc analysis of the clusters.

\textbf{Internal validity} relates to any confounding factor that could influence the results of the study. In this work, internal validity threats can be due to the manual classification step of the bug-fix clusters. A first threat is that we inspect a sample of items in each cluster (i.e., we select five elements) to decide if the cluster is a bug-fix pattern or if it is another kind of code change (e.g., refactoring, bug-fix induced, etc.). To mitigate this threat, we select the group of bug-fix to inspect randomly to avoid any correlation with time. A second threat is due to the manual classification that could potentially bias the results. To reduce the risk of this threat, each bug-fix pattern is independently classified by three authors, and combined through majority voting.

\textbf{External validity} relates to the possibility of generalizing the results of the study. This study focused on the three major OpenStack projects (i.e., Nova, Neutron, and Cinder). Even if our methodology is applicable to other projects, the bug-fixing patterns we found do not necessarily apply to other projects. However, the three projects we consider are large and diverse enough to get interesting insights on the similarity of bug-fixing patterns across different projects and across different languages (e.g., Python versus C and Java), and on the relevance of the context features. The diversity of the projects was reflected by differences in terms of project-specific patterns, due to the programming idioms, API conventions, and QA process of the projects, and in terms of the different context features that are relevant for the bug-fix patterns. This diversity allowed us to draw observations on the variability of patterns and on the relevance of context features in three large Python projects.

\textbf{Reliability validity} relates to the possibility of replicating this study. To ease replication of this study, we published the whole dataset with all of the features (code change and context), along with the \textit{PySA} tool. Moreover, in \sectionautorefname~\ref{sec:methodology} we provided detailed information on the methodological steps, algorithms and software involved, and choice of parameters.

\vspace{-0.5mm}
\section{Conclusion}
\label{sec:conclusion}
\vspace{-0.1mm}

In this paper, we propose an approach for analyzing bug-fixing changes, not limiting to \emph{what} has been changed, but also considering \emph{where} the change was made. We analyze bug-fixing changes by using a clustering approach on a set of features on the code change, in order to identify recurrent patterns. Furthermore, we investigate the \emph{context} of the bug-fix by analyzing an additional set of features derived from the code that surrounds the code change.

We applied the methodology to analyze bug-fixing changes in the OpenStack cloud computing platform, which is one of the most complex and widespread Python project, as it is the basis for several commercial products and services. 
We found that in some cases the bug-fix patterns are consistent with previous studies made on Java and C software, but in many other cases the bug-fix patterns are influenced by the Python language. Additionally, some recurrent patterns are strictly related to the nature of the specific project. The analysis of where the change occurred pointed out that bug-fixes are in all cases located in specific source-code contexts. 

The results of this study are valuable for several software engineering tasks that rely on knowledge of recurrent characteristics of software bugs. For example, mutation and fault injection testing will benefit in terms of decreasing the search space for mutants and the number of potential locations for injecting faults. A future direction for this work is to leverage these results by incorporating them into software engineering techniques and tools, such as in the context of fault injection and mutation testing.

\section*{Acknowledgments}
This work has been partially supported by the PRIN project ``GAUSS'' funded by MIUR (n. 2015KWREMX\_002) and by UniNA and Compagnia di San Paolo in the frame of Programme STAR.

%

\IEEEtriggeratref{20}

\bibliographystyle{IEEEtran}
\bibliography{IEEEabrv,bibliography}

\onecolumn

\appendices
\label{appendix:related_table}

\begin{table*}[!htb]
    \caption*{APPENDIX A\\Research Studies in the Area of Bug-Fixing Changes}
    \label{table:related_work}
    \begin{adjustbox}{width=\columnwidth,center}

    \begin{tabular}{>{\centering\arraybackslash}m{1in} >{\centering\arraybackslash}m{1in} >{\centering\arraybackslash}m{.5in} >{\centering\arraybackslash}m{1in} >{\centering\arraybackslash}m{1in} >{\arraybackslash}m{3.4in}}
                 
    
    \rowcolor{Gray} \multicolumn{1}{>{\centering\arraybackslash}m{1in}}{\textbf{Reference}} 
    & \multicolumn{1}{>{\centering\arraybackslash}m{1in}}{\textbf{Purpose}} 
    & \multicolumn{1}{>{\centering\arraybackslash}m{.5in}}{\textbf{Language}}
    & \multicolumn{1}{>{\centering\arraybackslash}m{1in}}{\textbf{Number of projects}}
    & \multicolumn{1}{>{\centering\arraybackslash}m{1in}}{\textbf{Dataset}}
    & \multicolumn{1}{>{\centering\arraybackslash}m{3.4in}}{\textbf{Findings}}\\
    \hline

Tufano et al. \cite{tufano2018learning}       & Mutation Testing & Java & GitHub projects between March 2011 and October 2017 on GitHub Archive) & 10,056,052 bug-fixing commits & The generated mutants perfectly correspond to the original buggy code in 9\% to 45\% of cases (depending on the model). Furthermore, the specialized models are able to inject different types of mutants. Mostly, the type of mutants is related to \textit{deletion of method calls}, on \textit{deletion and replacement of an argument in a method call}, on \textit{if-else blocks and its logical conditions}, \textit{deleting and replacing variable assignments}. \\     

    \hline

Brown et al. \cite{brown2017care} & Mutation Testing & C & The top 50 project repositories in GitHub & $\sim$600,000 commits containing $\sim$20 million individual diff blocks spanning 850 million lines of text.               & The authors provide an approach to automatically harvesting mutation operators—wild-caught mutants and compare the capabilities of the harvested mutation operators to those of existing studies. The proposed approach produces novel mutation operators, in turn creating defects that are about as difficult to kill as those arising from existing synthetic mutation operators. For example, the authors found new mutation operators like the missing call to a one-argument function whose return type is equal to its argument's type, direct access of field, and specific literal replacements.\\

    \hline

Zhong et al. \cite{zhong2015empirical} & Automatic Program Repair & Java & 5 projects (Aries, Cassandra, Derby, Lucene/Solr, Mahout) & 9,000 real-world bug fixes & The authors found the most frequent \textit{actions} related to the bug-fix, focusing on the AST node type of JDT library. According to that, the top 3 actions (addition, deletion, and modification) belong to \textit{JavaDOC}, \textit{ExpressionStatement}, \textit{MethodDeclaration}, and \textit{ReturnStatement}. However, there is an open discussion whether changes on documentation should be considered as bug-fixes or not. Furthermore, the authors found that in most cases a bug-fix consists of multiple edit actions, thus fault localization tools could found only 1 bug precisely. \\

    \hline

Soto et al. \cite{soto2016deeper}             & Automatic Program Repair & Java          & 554,864 Java projects from 2015 September Github repository offered by Boa & 4,590,679 bug fixing commits & The most common pattern observed is \textit{ABC (add or remove a branch condition)}; and the least common pattern is \textit{AOB (adding an array out of bound checker)}. If we conservatively assume that these patterns never appear together, they cover 14.78\% of buggy files. \\

    \hline
Koyuncu et al. \cite{koyuncu2018fixminer}        & Automatic Program Repair & Java          & 50 large and popular open-source projects & 8,009 patches & The top 5 clusters found are: (i) \textit{Method reference modification}, (ii) \textit{Variable declaration statement modification}, (iii) \textit{String value modification in method call}, (iv) \textit{Method call parameter modification}, (v) \textit{Constant modification in declaration statement}. 
Furthermore, in the 80\% of the cases FixMiner generates patches that are correct to be used in APR task. The closest related works \cite{le2016history, jiang2018shaping}, achieve respectively 26\% and 70\% of correctness.\\

    \hline
Lin et al. \cite{lin2016empirical}           & Bug characterization     & Python        & 10 python projects (Django, Tornado, Pandas, Pylearn2, Numpy, Scipy, Sympy, Nltk, Beets, Mopidy)                                                                                                           & 132,294 commits                                                                                                            & In most projects studied, \textit{Function Change} and \textit{Statement Change} are the most common change types. \textit{Loop Structure Change} is the most uncommon change type. The distributions of change type frequency share similar trends across studied projects. There are no significant differences among the distributions of change type frequency across studied domains. In the studied projects, if structure related change types are more related to bug-fix, especially \textit{Conditional Expression Update} and \textit{If Insert}. \\

    \hline

Musavi et al. \cite{musavi2016experience}        & Bug characterization     & Python        & Openstack project (the Nova, Swift, Heat, Neutron and Keystone projects)                                                                                                                                   & 221,671 commits from 2010-05 to 2015-02                                                                                    & The authors found that in the 56\% cases the cause of API failures is due to \textit{Small programming faults}, i.e., trivial programming mistake (e.g., the developer changes the default value of a variable to another value). The next most common type of fault (14\%) is \textit{major programming faults}. \textit{Configuration faults} (14\%). \\

    \hline
    
Osman et al. \cite{osman2014mining}       & Bug characterization     & Java          & 717 Java projects & 190,821 code changes corresponding to 94,534 bug-fix commits & In the 53\% of case, bug-fixes involve only one line of code. Specifically for bug characteristics: (i) More than 48\% of bugs are about \textit{missing NULL checks}; (ii) Other most frequent bug are \textit{Missing Invocation Method} and \textit{Wrong Parameters/Method}. \\

    \hline
    
Pan et al. \cite{pan2009toward} & Bug characterization & Java & 7 Java projects (ArgoUML, Columba, Eclipse, JEdit, Scarab, Lucene, and MegaMek) & 20,270 number of revisions, within 6,978 number of commits & In that study, the authors found 27 bug fix patterns, which include \textit{If-related (IF)}, \textit{Method Call (MC)}, \textit{Loop (LP)}, \textit{Assignment (AS)}, \textit{Switch (SW)}, \textit{Try (TY)}, \textit{Method Declaration (MD)}, \textit{Sequence (SQ)}, and \textit{Class Field (CF)}. 
The most common categories of bug fix patterns are \textit{Method Call} (MC, 21.9–33.1\%) and \textit{If-Related} (IF, 19.7–33.9\%). 
The most common individual patterns are \textit{MC-DAP (method call with different actual parameter values)} at 14.9–25.5\%, \textit{IF-CC (change in if conditional)} at 5.6–18.6\%, and \textit{AS-CE (change of assignment expression)} at 6.0–14.2\%. \\

    \hline

Martinez et al. \cite{martinez2013automatically} & Bug characterization     & Java          & 6 projects                                                                                                                                                                                                 & 33,365 revisions, 6,233 commits                                                                                            & For instance, adding new methods (MD-ADD) and changing a condition expression (IF-CC) are the most frequent patterns while adding a try statement (TY-ARTC) is a low frequency action for fixing bugs. \\

    \hline
    
Fluri et al. \cite{fluri2008discovering}         & Bug characterization     & Java          & 3 projects (jEdit, JFreeChart, and Webframework (a commercial Java framework for web applications))                                                                                                        & 30,930 revisions with 229,604 changes & The authors found 2 top clusters for \textit{if-statement} and \textit{throw statement inserts} for JEDit e JFreeChart projects. About  WebFramework project, the top clusters are about \textit{Constructor invocation changes}, \textit{Return type based method renaming}, \textit{Introducing prefixed parameter names}, \textit{Introducing single exit}, \textit{Change existing exception handling}. The authors do not provide any quantitative information for the patterns found. \\

    \hline

Duraes et al. \cite{duraes2006emulation}         & Fault Injection          & C             & 12 projects (CDEX, Vim, FreeCiv, pdf2h, GAIM, Joe, ZSNES, Bash, Linux Kernel, Firebird, MingW, ScummVM)                                                                                                    & 668 bugs                                                                                                                    & According to the ODC classification, the authors found that: (i) \textit{Algorithm class} are the dominant faults (40.1\%). In particular, the 2 most frequent are about \textit{Missing IF construct plus statements} (30\%) and \textit{Missing Function Call} (26\%); (ii) \textit{Assignment faults} have approximately the same weight as \textit{Checking faults} (21.4\% and 25\%); (iii) \textit{Interface} and \textit{Function faults} are the less frequent ones (7.3\% and 6.1\%). \\

    \hline
        
Basso et al. \cite{basso2009investigation}       & Fault Injection          & Java          & 6 projects (Azureus Vuze, FreeMind, JEdit, Phex, Struts, Tomcat)                                                                                                                                           & 574 bugs                                                                                                                  & According to ODC classification, the 2 most frequent fault type are \textit{Missing Functionality} and \textit{Missing if construct plus statements} (30\%). The third most frequent (10.5\%) is \textit{Missing Function Call fault}. \\

    \hline

Neamtiu et al. \cite{neamtiu2005understanding}   & Refactoring              & C             & 5 projects (Apache, OpenSSH, Vsftpd, Bind, and the Linux kernel)                                                                                                                                           & N/A                                                                                                                        & The authors found that: \textit{(i)} the function and global variable \textit{additions} are far more frequent than \textit{deletions}; \textit{(ii)} the rates of \textit{addition} and \textit{deletion} vary from program to program; \textit{(iii)} the function bodies change quite frequently over time, but function prototypes change only rarely; \textit{(iv)} the type definitions (like \textit{struct} and \textit{union} declarations) change infrequently, and often in simple ways. \\

    \hline
Silva et al.\cite{silva2017refdiff} & Refactoring & Java & 7 projects & N/A & The authors propose a tool for detecting 12 well-known refactoring types \cite{prete2010template}. The proposed approach achieved the best result among the evaluated tools in the state-of-the-art, with a \textit{Precision} of 1.00 and \textit{Recall} of 0.88.\\

        \hline
Hora et al. \cite{hora2018assessing}             & Refactoring              & Java          & 15 large projects in Java & The commits range from 1,025 (Android Image Loader) to 39,389 (Kotlin)                                                     & The most frequent untracked changes happen at the method level and are due to Rename Method (26\%), Extract Method (23\%), and Move Method (22\%). In contrast, the least frequent ones are due to Extract Superclass ($<$1\%), Extract Interface (1\%), and Push Down Method (1\%). The ratio of untracked changes ranges from 10\% to 21\% for methods, and from 2\% to 15\% for classes. Thus, the threat is more frequent at the method level. \\

    \hline
    




\end{tabular}
\end{adjustbox}
\end{table*}

\end{document}